\documentclass[journal]{IEEEtran}
\ifCLASSINFOpdf
\else
\fi
\usepackage{bm}
\usepackage{indentfirst}
\usepackage{epsfig}
\usepackage{amsfonts}
\usepackage{amsmath}
\usepackage{url}
\usepackage{color}
\usepackage{algorithm}
\usepackage{algpseudocode}
\usepackage{cancel}
\usepackage{mathcomp}
\usepackage{footnote}
\usepackage{cite}
\usepackage{mathtools}
\usepackage{amssymb}
\usepackage{amsmath}
\usepackage{amsthm}
\newtheorem{theorem}{Theorem}

\newtheorem{proposition}{Proposition}
\usepackage{mathrsfs}
\usepackage{amsfonts}
\usepackage{graphicx}
\usepackage[justification=centering]{caption}
\usepackage{subcaption}
\usepackage{multirow}
\usepackage{hhline}
\usepackage{bigstrut}
\usepackage{cases}
\usepackage{verbatim}
\usepackage[utf8x]{inputenc}
\usepackage[T1]{fontenc}
\usepackage{xcolor}
\usepackage{epstopdf}

\begin{document}
	\title{Incentivizing Energy Trading for Interconnected Microgrids}

	\author{Hao~Wang,~\IEEEmembership{Member,~IEEE,}
		and~Jianwei~Huang,~\IEEEmembership{Fellow,~IEEE}

		\thanks{This work was supported by the Theme-based Research Scheme (Project No. T23-407/13-N) from the Research Grants Council of the Hong Kong Special Administrative Region, China, and a grant from the Vice-Chancellor's One-off Discretionary Fund of The Chinese University of Hong Kong (Project No. VCF2014016), and in part by the State of Washington through the University of Washington Clean Energy Institute. Part of the results have appeared in IEEE ICC 2015 \cite{ourpaper}.}
		\thanks{H. Wang is with Clean Energy Institute, University of Washington, Seattle, WA 98195 USA (e-mail: hwang16@uw.edu).}
		\thanks{J. Huang is with the Network Communications and Economics Lab (NCEL), Department of Information Engineering, The Chinese University of Hong Kong, Shatin, Hong Kong SAR, China (e-mail: jwhuang@ie.cuhk.edu.hk).}
	}
	\maketitle

	\begin{abstract}
		In this paper, we study the interactions among interconnected autonomous microgrids, and propose a joint energy trading and scheduling strategy. Each interconnected microgrid not only schedules its local power supply and demand, but also trades energy with other microgrids in a distribution network. Specifically, microgrids with excessive renewable generations can trade with other microgrids in deficit of power supplies for mutual benefits. Since interconnected microgrids operate autonomously, they aim to optimize their own performance and expect to gain benefits through energy trading. We design an incentive mechanism using Nash bargaining theory to encourage proactive energy trading and fair benefit sharing. We solve the bargaining problem by decomposing it into two sequential problems on social cost minimization and trading benefit sharing, respectively. For practical implementation, we propose a decentralized solution method with minimum information exchange overhead. Numerical studies based on realistic data demonstrate that the total cost of the interconnected-microgrids operation can be reduced by up to 13.2\% through energy trading, and an individual participating microgrid can achieve up to 29.4\% reduction in its cost through energy trading.
	\end{abstract}
	
	\begin{IEEEkeywords}
		Microgrid, renewable energy, energy storage, demand response, energy trading, Nash bargaining solution (NBS), alternating direction method of multipliers (ADMM), distributed algorithm.
	\end{IEEEkeywords}

	\section{Introduction}
	Traditional power systems often generate power in large power stations using fossil fuel resources, and distribute it over long distances. This results in quick depletion of fossil fuel resources, increased environmental pollution, and potentially significant energy losses during transmission and distribution. This motivates the study and adoption of microgrids \cite{mgtrend,mgs}, which are small-scale power supply networks with distributed generations and demands. A microgrid consists of an interconnected network of several energy sources (including both conventional and renewable energy generators), and serves the local electrical loads from residential, commercial, and industrial consumers. In comparison with centralized and conventional models of power system, the microgrid brings several benefits: reducing power transmission loss, enhancing the system resilience, and integrating distributed generations especially from renewable sources. Wide deployment and proper management of microgrids can have significant positive impact on the overall power grid system. 
	
	Several recent representative studies on the interactions among multiple microgrids include \cite{mg5,mg6,mg8,mg9} and the references therein. However, the studies in \cite{mg5,mg6,mg8,mg9} either assumed that all the microgrids are coordinated by a common grid operator, or focused on the interaction between the main grid and microgrids in a hierarchical structure. Although these scenarios are practically important, it is equally important to consider the scenario that involves multiple small autonomous microgrids operating in an distributed fashion, which is considered as an important feature of the next generation smart grid \cite{mgtrend,mgs}. These interconnected microgrids can exchange energy and information with each other, and are operated by independent microgrid operators instead of a common coordinator. This essentially leads to an (small scale) energy market of interconnected microgrids. Thus, it is important to design a new operation framework for this new decentralized paradigm.	
	
	To design a new distributed optimization operation framework for interconnected microgrids, we need to tackle the following two challenges: \emph{incentive issues} and \emph{decision coupling}. First, autonomous microgrids are independent entities with self-interests, and they will only interact with other microgrids if such interactions lead to additional benefits. Hence it is essential to design incentive mechanisms that encourage such interactions. Second, each microgrid has two different types of operating decisions: \emph{external} strategy on how to interact with other microgrids and \emph{internal} strategy on how to coordinate local power supply and demand. These two types of operating decisions are coupled, \emph{e.g.}, selling energy to other microgrids will reduce the energy supply that can be used to satisfy local needs. Similarly, the decisions of different microgrids are also tightly coupled together, as the total supply and demand in the market depend on every microgrid's decision. 
	
	In this paper, we propose an incentive mechanism using Nash bargaining solution, to encourage proactive interactions and fair benefit sharing among interconnected microgrids. All the interconnected microgrids jointly optimize their energy trading and scheduling, by taking the advantages of diverse supply and demand patterns in different microgrids. The key idea is to exploit the fact that supply and demand profiles in different microgrids exhibit both time and location diversities. Due to the time-varying and location-dependent nature of renewable energy generations, one microgrid may have excess local renewable generation at the same time when another microgrid is deficient in power supply from its local generation. In addition, users' power consumption profiles in different microgrids can also be significantly different, because of various types of consumers. Measurement data show that the residential users consume more power in the night, while the commercial power usage reaches peak during day time \cite{loadprofile}. The diversified renewable outputs and demand profiles provide ample opportunities for interconnected microgrids to exchange electricity with each other to enhance their operational performance and reduce operating cost. 
	
	The main contributions of this paper are listed as follows.
	
	\begin{itemize}
		\item \textit{Joint scheduling and trading}: We develop a holistic model for the microgrids-system to jointly optimize power scheduling within individual microgrids and energy trading among interconnected microgrids.
		
		\item \textit{Incentive mechanism design}: We propose a bargaining-based incentive mechanism for energy trading among interconnected microgrids, which can leverage their diverse supply/demand profiles and bring mutual benefits.
		
		\item \textit{Problem decomposition and distributed solution}: We decompose the bargaining problem into two sequential subproblems to solve the optimal energy schedules and trading payments, respectively. We also design a distributed solution method with limited information exchange overhead that is suitable for the practical implementation.
		
		\item \textit{Numerical simulations and implications}: Numerical studies based on realistic data demonstrate the effectiveness of the energy trading solution, with a total cost reduction of 13.2\% for the interconnected-microgrids system.
	\end{itemize}
	
	The remainder of this paper is organized as follows. We review the related work in Section II. In Section III, we formulate the joint energy trading and scheduling of interconnected microgrids as a Nash bargaining problem. We present theoretical analysis of the problem and propose a decentralized solution method in Section IV and V, respectively. Numerical results are presented in Section VI. Finally, we conclude this paper in Section VII.

	\section{Related Work}
	Operation of the multiple-microgrids system has attracted extensive research in \cite{mg5,mg6,mg8,mg9}. Fathi and Bevrani studied cooperative power dispatching of multiple interconnected microgrids in \cite{mg5}, and considered the impact of demand uncertainty in \cite{mg6}. The studies in \cite{mg5,mg6} all assumed that the microgrids are coordinated by a common operator. This may not always be the case in practice, where microgrids can be self-managed and independent entities. Asimakopoulou \emph{et al.} \cite{mg8} proposed a leader-follower energy management strategy to study the interactions between an energy producer and energy service providers. Wang \emph{et al.} \cite{mg9} studied the interactions between a distribution network operator and clusters of microgrids. The studies in \cite{mg8,mg9} have mainly focused on the interactions between the main grid and microgrids under a hierarchical structure.
	
	Several other studies explored direct interactions among interconnected microgrids \cite{mg14,mg15}. Matamoros \emph{et al.} \cite{mg14} studied energy trading between two islanding microgrids. Gregoratti and Matamoros \cite{mg15} explored energy exchange among multiple microgrids, and formulated a convex optimization problem to minimize the global cost. However, studies in \cite{mg14} and \cite{mg15} did not consider the self-interests of multiple microgrids. Recent work \cite{mg10,addgame2,addagent4,addagent5} studied the energy trading and management problems of microgrids using game-theoretic approaches. For example, Zhang \emph{et al.} in \cite{mg10} proposed a randomized auction framework for microgrids to participate in the electricity market. Wang \emph{et al.} in \cite{addgame2} proposed a double-auction market for distributed storage units to trade energy in the smart grid. Nunna \emph{et al.} proposed the agent-based energy management methods to facilitate energy trading among microgrids with demand response in \cite{addagent4} and with distributed storage in \cite{addagent5}. In this paper, we aim to study the energy trading among multiple interconnected microgrids, considering the self-interest and diverse generation and load profiles of each microgrid. We formulate a holistic model for the interconnected-microgrids system and jointly optimize both their internal power scheduling and external energy trading. We design an incentive mechanism to encourage proactive energy trading among interconnected microgrids, such that each participating microgrid can benefit from the trading. We propose a two-step solution method, and design a distributed algorithm to solve energy trading and payment. Finally, we validate the performance of our proposed method using realistic data.

	\section{Interconnected-Microgrids System}
\begin{figure}[t] 
		\centering
		\includegraphics[width=6.5cm]{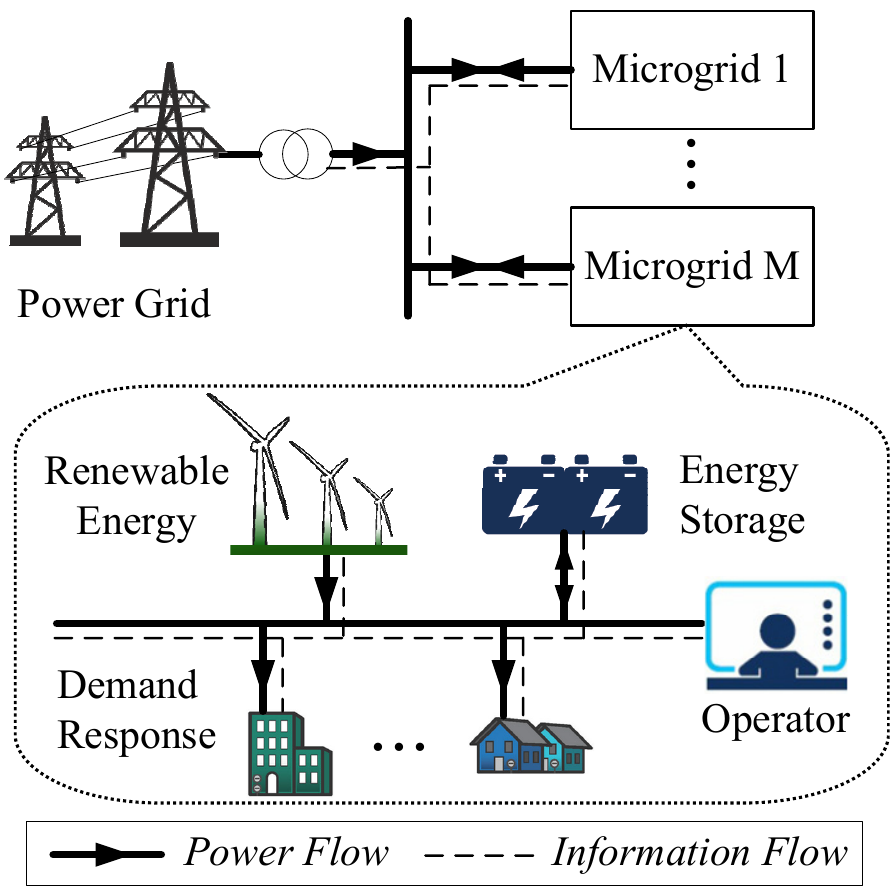}
		\caption{\label{fig-system} A system with interconnected microgrids.}
	\end{figure}

	We consider a network of $M$ interconnected microgrids $\mathcal{M}=\{1,...,M \}$. These microgrids are connected to the main power grid\footnote{Microgrid can work either in the island mode or in the connected mode. In the island mode, the microgrid is isolated from the main grid, and only relies on its local generation as the supply. In the connected mode, the microgrid is connected to the main grid, and can purchase power from the main grid. We assume that the microgrid works in the connected mode in this paper.}, and are also interconnected with each other. These interconnected microgrids can exchange power and information with each other through a power bus and a communication network. Fig. \ref{fig-system} illustrates such a system model. Each microgrid $i \in \mathcal{M}$ contains the following components: local renewable generation, energy storage, and demand responsive users. We consider an operation horizon of one day, which is divided into $T=24$ equal time slots, denoted as $\mathcal{T}=\{1,...,T\}$. We assume that microgrids can schedule their power supplies and flexible loads based on the prediction of daily renewable energy profiles, and any mismatch caused by the prediction errors can be balanced in the real-time market. In this paper, we focus on the energy trading and scheduling in a day-ahead market. The microgrid operator is responsible for the power scheduling in the microgrid as well as its energy trading with other interconnected microgrids at the beginning of each day. 
	
	As power scheduling and energy trading are highly coupled across microgrids, we aim at the joint optimization of all the interconnected microgrids in a distributed fashion. Before presenting the interconnected energy trading model, we formulate the power scheduling problem within each microgrid.
	
	\subsection{Power Scheduling in Each Microgrid}
	Power supply in microgrid $i$ can be categorized into local renewable power generation and power from the main grid.
	
	\subsubsection{Local renewable power generation}
	There are various types of renewable energy technologies, such as wind, photovoltaic, biomass, and tidal systems. We will focus on the renewable power generation from the wind source as a concrete example in this paper,\footnote{Our theoretic model and analysis are applicable to other renewable sources.} as our later numerical evaluations will be based on the wind speed data obtained from the Hong Kong Observatory \cite{hkdata}.
	
	Based on the hourly wind speed data \cite{hkdata}, we use a wind power model \cite{windmodel} to calculate the corresponding wind power generation $\boldsymbol{\eta}_{i} = \{ \eta_{i}^{t},~\forall t \in \mathcal{T} \}$ in microgrid $i$ per $kW$ capacity. We assume that microgrid $i$ has installed wind turbine generators with total generation capacity $G_i$ ($kW$), and have the following constraint for the wind power supply $\boldsymbol{g}_{i} = \{ g_{i}^{t},~\forall t \in \mathcal{T} \}$: 
	\begin{align}
		0 \leq g_{i}^{t} \leq \eta_{i}^{t} G_{i}, \;\forall t\in\mathcal{T}, \;\forall i\in\mathcal{M}, \label{constraint-wind}
	\end{align}
	where $\eta_{i}^{t} G_{i}$ denotes the maximum available wind power of microgrid $i$ in time slot $t$. Note that the local wind power usage $g_{i}^{t}$ of microgrid $i$ in time slot $t$ can be less but no greater than the available wind power $\eta_{i}^{t} G_{i}$, because the microgrid operator can either curtail excess wind power generation or sell excess generation back to the main grid.
	
	Different from conventional power generations, wind power generation does not consume fuel sources, so for simplicity we assume a zero marginal generation cost \cite{zerocost}.
	
	\subsubsection{Main grid power}
	When the local wind power generation is not adequate to meet the demand, microgrid $i$ can also purchase power from the main power grid. Let $q_{b,i}^{t}$ denote the power bought from the main grid by microgrid $i$ in time slot $t$, and define $\boldsymbol{q}_{b,i} =\{ q_{b,i}^{t},~\forall t \in \mathcal{T} \}$. The purchased power $\boldsymbol{q}_{b,i}$ is subject to the following constraint:
	\begin{align}
		0 \leq q_{b,i}^{t} \leq Q_{b,i}^{ \max}, \;\forall t\in\mathcal{T}, \;\forall i\in\mathcal{M}, \label{constraint-purchased}
	\end{align}
	where $Q_{b,i}^{ \max}$ denotes the maximum amount of power that microgrid $i$ can purchase from the main grid, due to physical capacity limit.
	
	When the microgrid has excess local wind power generation, it can sell power back to the main grid under a feed-in tariff contract \cite{feedin}. Let $q_{s,i}^{t}$ denote the power sold to the main grid by microgrid $i$ in time slot $t$, and define $\boldsymbol{q}_{s,i} =\{ q_{s,i}^{t},~\forall t \in \mathcal{T} \}$. The power sold to the main grid $\boldsymbol{q}_{s,i}$ satisfies the following constraint:
	\begin{align}
	    0 \leq q_{s,i}^{t} \leq Q_{s,i}^{ \max}, \;\forall t\in\mathcal{T}, \;\forall i\in\mathcal{M}, \label{constraint-sold}
	\end{align}
	where $Q_{s,i}^{ \max}$ denotes the maximum allowed amount of power that microgrid $i$ can sell to the main grid.
	
	Energy cost is associated with the main grid power schedules and prices. We denote $p_{b}^t$ as the power procurement (buying) price and denote $p_{s}^t$ as the power feed-in (selling) price\footnote{We assume that all the microgrids are connected to the same main grid, and the power buying/selling prices for all the microgrids are the same.} set by the main grid in time slot $t$. For notational simplicity, we let $\boldsymbol{q}_{i} = \{ \boldsymbol{q}_{b,i},\boldsymbol{q}_{s,i} \}$ denote the main grid power schedule of microgrid $i$ over $T$ time slots. The energy cost of microgrid $i$ is written as
	\begin{align}
		C_{i} (\boldsymbol{q}_{i})= \sum_{t\in\mathcal{T}} \left( p_{b}^{t} q_{b,i}^{t} - p_{s}^{t} q_{s,i}^{t} \right), \;\forall i \in \mathcal{M}. \label{cost-purchased}
	\end{align}

	\subsubsection{Local power demand}
	Let $\mathcal{N}_i$ denote the set of users in microgrid $i \in \mathcal{M}$. We classify the loads of each user $n \in \mathcal{N}_i$ into two categories: inelastic loads and elastic loads. 
	
	The inelastic loads, such as refrigerator and illumination demands, cannot be easily shifted over time. We let $b_{i}^{t}$ denote the aggregate inelastic load of all the users in microgrid $i$ and time slot $t$, and denote $\boldsymbol{b}_{i} = \{ b_{i}^{t},~\forall t \in \mathcal{T} \}$. The elastic loads, such as electric vehicle, HVAC (heating, ventilation and air conditioning) and washer demands, can be flexibly scheduled over time. For user $n \in \mathcal{N}_i$ in microgrid $i$, the elastic load is denoted as $\boldsymbol{x}_{n} = \{ x_{n}^{t} ,~\forall t \in \mathcal{T} \}$, where $x_{n}^{t}$ is user $n$'s elastic power consumption in time slot $t$.
	
	The demand response program can only control the elastic loads, and should be subject to the following constraints:
	\begin{align}
		& \sum_{t \in \mathcal{T}} x_{n}^{t} = D_{n},~\forall n \in \mathcal{N}_i,~\forall i \in \mathcal{M}, \label{constraint-load1} \\
		& d_{n}^{t, \min} \leq x_{n}^{t} \leq d_{n}^{t, \max},\;\forall t\in\mathcal{T},~\forall n \in \mathcal{N}_i,~\forall i \in \mathcal{M}, \label{constraint-load2} 
	\end{align}
	where constraint (\ref{constraint-load1}) corresponds to the prescribed total energy requirement $D_{n}$ in the entire operation horizon. Constraint (\ref{constraint-load2}) provides a lower bound $d_{n}^{t, \min}$ and upper bound $d_{n}^{t, \max}$ for the power consumption of user $n$ in each time slot $t$.
	
	Elastic power consumption of user $n$ can be scheduled across time as long as the power consumption satisfies the constraints (\ref{constraint-load1}) and (\ref{constraint-load2}). However, scheduling power load may affect user's comfort. Let $\boldsymbol{y}_{n} = \{ y_{n}^{t} ,~\forall t \in \mathcal{T} \}$ denote the most preferred power consumption of user $n$. When the actual power consumption $\boldsymbol{x}_{n}$ deviates from the preferred power consumption $\boldsymbol{y}_{n}$, user $n$ will experience discomfort. Similar as the discomfort measure in \cite{discomfort}, we define the discomfort cost of user $n$ as
	\begin{align}
		& C_{n}(\boldsymbol{x}_{n}) = \beta_{n} \sum_{t \in \mathcal{T}} \left( x_{n}^{t}-y_{n}^{t} \right)^{2} , \label{cost-load}
	\end{align}
	where $\left( x_{n}^{t}-y_{n}^{t} \right)^{2}$ measures how much the actual power consumption differs from the preferred power consumption. Weighted coefficient $\beta_{n}$ is used to indicate the sensitivity of user $n$ towards the power consumption deviation.

	\subsubsection{Local energy storage}
	Energy storage (such as batteries) can smooth out the intermittent wind power generation, flatten the power load by charging when the load is low and discharging during peak load times, and exploit time-varying electricity prices for arbitrage. We assume that microgrid $i$ has installed energy storage devices with a total capacity $S_{i}^{ \max}$, and let $s_{i}^{t}$, $r_{c,i}^{t}$, and $r_{d,i}^{t}$ denote the amount of electricity stored, charged, and discharged in time slot $t$, respectively.
	
	First, the charging and discharging power in each time slot $t$ are bounded, and satisfy the following constraints:
	\begin{align}
		& 0 \leq r_{c,i}^{t} \leq r_{c,i}^{ \max},~\forall t\in\mathcal{T}, \;\forall i\in\mathcal{M}, \label{constraint-storage1} \\
		& 0 \leq r_{d,i}^{t} \leq r_{d,i}^{ \max},~\forall t\in\mathcal{T}, \;\forall i\in\mathcal{M}, \label{constraint-storage2} 
	\end{align}
	where $r_{c,i}^{ \max}$ and $r_{d,i}^{ \max}$ denote the maximum charging and discharging rates of energy storage in microgrid $i$, respectively.
	
	Second, there are power losses when electricity is charged into and discharged from the battery. We denote $\eta_{c,i} \in \left( 0,1 \right]$ and $\eta_{d,i} \in \left( 0,1 \right]$ as the conversion efficiencies of charging and discharging. Therefore, we obtain the energy storage dynamics of microgrid $i$ in time slot $t$ as
	\begin{align}
		& s_{i}^{t} = s_{i}^{t-1} + \eta_{c,i} r_{c,i}^{t} - \frac{r_{d,i}^{t}}{\eta_{d,i}}, ~\forall t \in \mathcal{T}, \; \forall i \in \mathcal{M}. \label{constraint-storage3} 
	\end{align}
	
	Third, repeated charging and discharging cause degradation of the energy storage devices. The life-time of energy storage is usually characterized by the number of charging/discharging cycles under a given depth of discharge (DoD) \cite{DOD}, which is defined as the maximum discharge to the capacity. The storage can sustain more charging/discharging cycles with smaller DoD. We denote $DoD_{i}$ as the DoD requirement for energy storage operation in microgrid $i$, and have the following constraint for the energy level:
	\begin{align}
		& (1 - DoD_{i}) S_{i}^{ \max} \leq s_{i}^{t} \leq S_{i}^{ \max}, ~\forall t \in \mathcal{T}, \;\forall i \in \mathcal{M},\label{constraint-storage4}  
	\end{align}
	where $S_{i}^{ \max}$ and $(1 - DoD_{i}) S_{i}^{ \max}$ are upper and lower bounds for the level of stored energy in microgrid $i$, respectively. Specifically, the stored energy should be no greater than the physical capacity $S_{i}^{ \max}$, and no less than the operational requirement $(1 - DoD_{i}) S_{i}^{ \max}$. We also restrict the terminal energy level $s_{i}^{T}$ to be the same as its initial level $s_{i}^{0}$, such that the daily storage operation is decoupled across different days.
	
	Last, to incorporate the degradation of energy storage caused by charging and discharging, we introduce $c_{s}$ as the amortized cost of charging and discharging over the lifetime, and model the cost of energy storage operation \cite{storagecost} as
	\begin{align}
		C_{s} (\boldsymbol{r}_{c,i},\boldsymbol{r}_{d,i}) = c_{s} \left( \sum_{t\in\mathcal{T}} r_{c,i}^{t} + \sum_{t\in\mathcal{T}} r_{d,i}^{t} \right), \label{cost-storage}
	\end{align}
	where $\boldsymbol{r}_{c,i} = \{ r_{c,i}^{t},~\forall t\in\mathcal{T} \}$ and $\boldsymbol{r}_{d,i} = \{ r_{d,i}^{t},~\forall t\in\mathcal{T} \}$ denote the charging and discharging amount over the operation horizon $\mathcal{T}$ in microgrid $i$, respectively.

	\subsection{Single Microgrid's Cost Minimization Problem}
	In each microgrid, its operator coordinates the power scheduling, energy storage charging, discharging, and elastic load shifting. First of all, the microgrid operator should keep the power supply and demand balanced, \emph{i.e.},
	\begin{align}
		g_{i}^{t} + q_{b,i}^{t} + r_{d,i}^{t} = q_{s,i}^{t} + r_{c,i}^{t} + b_{i}^{t} + \sum_{n \in \mathcal{N}_i} x_{n}^{t},\; \forall t\in\mathcal{T}, \forall i\in\mathcal{M}, \label{constraint-balance}
	\end{align}
	where the left-hand side represents the total power supply in time slot $t$, including wind power generation $g_{i}^{t}$, power drawn from the main grid $q_{b,i}^{t}$, and power discharged from the battery $r_{d,i}^{t}$. The right-hand side represents the total power demand in time slot $t$, including power sold to the main grid $q_{s,i}^{t}$, power charged into the battery $r_{c,i}^{t}$, aggregate inelastic load $b_{i}^{t}$ and aggregate elastic load of all the users $\sum_{n \in \mathcal{N}_i} x_{n}^{t}$.
	
	Moreover, the power sold to the main grid $q_{s,i}^{t}$ cannot be greater than the total available power of microgrid $i$, \emph{i.e.},
	\begin{align}
		q_{s,i}^{t} \leq \eta_{i}^{t} G_{i} - g_{i}^{t} + s_{i}^{t},\; \forall t\in\mathcal{T}, \forall i\in\mathcal{M}, \label{constraint-selling}
	\end{align}
	where the total available power in time slot $t$ consists of the local wind power surplus $\eta_{i}^{t} G_{i} - g_{i}^{t}$ and the battery energy level $s_{i}^{t}$.
	
	We assume that the microgrid operator owns both renewable energy generators and energy storage facilities, and can schedule the elastic loads through demand response programs. The objective of each microgrid operator is to minimize its total operating cost, including the energy cost, energy storage operation cost, and users' discomfort costs. For simplicity, we denote the operating cost of microgrid $i$ as
	\begin{align*}
		& \displaystyle C_i^{O}( \boldsymbol{q}_{i}, \boldsymbol{x}_{n}, \boldsymbol{r}_{c,i}, \boldsymbol{r}_{d,i}) \\
		\triangleq & C_{i}(\boldsymbol{q}_{i}) + \sum_{n \in \mathcal{N}_i} C_{n}(\boldsymbol{x}_{n}) + C_{s}(\boldsymbol{r}_{c,i},\boldsymbol{r}_{d,i}).
	\end{align*}
	
	Therefore, we formulate the microgrid operator's operating cost minimization problem as follows\\
	
	\textbf{Cost minimization problem for Microgrid $i$ (P-MG$_i$)}
	\begin{equation*}
		\begin{aligned}
			& \min
			& & C_i^{O}( \boldsymbol{q}_{i}, \boldsymbol{x}_{n}, \boldsymbol{r}_{c,i}, \boldsymbol{r}_{d,i}) \\
			& \text{subject to}
			& & \text{\eqref{constraint-wind}--\eqref{constraint-sold}, \eqref{constraint-load1}, \eqref{constraint-load2}, \eqref{constraint-storage1}--\eqref{constraint-storage4}, \eqref{constraint-balance}, and \eqref{constraint-selling}} \\
			& \text{variables:}
			& & \boldsymbol{g}_{i},\boldsymbol{q}_{i},\boldsymbol{x}_{n},\boldsymbol{r}_{c,i},\boldsymbol{r}_{d,i}.
		\end{aligned}
	\end{equation*}
	
	We can verify that Problem \textbf{P-MG$_i$} is convex, and can be efficiently solved by several standard optimization techniques \cite{convex}. We let $C_{i}^{Non}$ denote the optimal value of the objective function in Problem \textbf{P-MG$_i$}, which indicates the minimum cost that microgrid $i$ can achieve without trading energy with other microgrids. It also serves as the \emph{noncooperative benchmark} for comparison in Section IV.

	\subsection{Energy Trading among Microgrids}
	Now we consider the possibility of energy trading among interconnected microgrids. Microgrids at different locations have different renewable power generations and local load profiles. Through trading energy with each other, interconnected microgrids can exploit the diversities of supply and demand patterns, and achieve mutual benefits. Next we will study the energy trading interactions among interconnected microgrids based on the Nash bargaining solution \cite{nash}. 
	
    A Nash bargaining problem solves a fair Pareto optimal solution and leads to a bargaining solution, which fulfills the following axioms \cite{nash}.
		\begin{enumerate}
			\item \emph{Individual Rationality}: All players should improve their utilities through the bargaining compared with their performances without cooperation (namely disagreement points); otherwise, they would not cooperate.
			\item \emph{Feasibility}: For the bargaining game, there exists at least one feasible solution that satisfies all the constraints.
			\item \emph{Pareto Optimality}: A player cannot find another solution, in which every player receives a utility no smaller than the one received in the Nash bargaining game, and some player receives a payoff that is strictly higher than the one received in the Nash bargaining game. 
			\item \emph{Independence of Irrelevant Alternatives}: If the bargaining solution is found on a smaller domain of the feasible set, then the solution is not affected by expanding the smaller domain within the feasible set. 
			\item \emph{Independence of Linear Transformations}: The bargaining solution is invariant if the utility function and disagreement point are scaled by a linear transformation.		 
			\item \emph{Symmetry}: If players have the same disagreement points and utility functions, they will have the same utility at the bargaining solution regardless of their indices.
		\end{enumerate}
	Axioms 1), 2), and 3) define the bargaining set, and axioms 4), 5), and 6) ensure the fairness of the bargaining solution.
	
	We consider that each microgrid $i \in \mathcal{M}$ bargains with all the other interconnected microgrids to determine the amount of energy trading $\boldsymbol{e}_{i} = \{ e_{i,j}^{t},~\forall t \in \mathcal{T},~\forall j \in \mathcal{M} \backslash i \}$ and the associated payment $\boldsymbol{\pi}_{i} = \{ \pi_{i,j},~\forall j \in \mathcal{M} \backslash i \}$. Here $e_{i,j}^{t}$ denotes the amount of energy that microgrid $i$ exchanges with microgrid $j$ in time slot $t$, and $\pi_{i,j}$ denotes the associated payment for energy trading between microgrid $i$ and microgrid $j$. If microgrid $i$ purchases energy from microgrid $j$ in time slot $t$, then $e_{i,j}^{t} > 0$; otherwise, microgrid $i$ sells energy to microgrid $j$ and $e_{i,j}^{t} < 0$. Similarly, if microgrid $i$ makes payment to microgrid $j$, then $\pi_{i,j} > 0$; otherwise microgrid $i$ receives payment from microgrid $j$ and $\pi_{i,j} < 0$.\footnote{Note that the payments $\boldsymbol{\pi}_{i}$ are determined through bargaining between microgrids, and are not necessary linear in terms of the energy exchange $\boldsymbol{e}_{i}$.}
	
	The energy trading and payment among microgrids should satisfy the market clearing constraints:
	\begin{align}
		& e_{i,j}^{t} + e_{j,i}^{t} = 0,~\forall t \in \mathcal{T},~\forall j \in \mathcal{M} \backslash i,~\forall i \in \mathcal{M}, \label{constraint-trading1} \\
		& \pi_{i,j} + \pi_{j,i} = 0,~\forall j \in \mathcal{M} \backslash i,~\forall i \in \mathcal{M}. \label{constraint-trading2} 
	\end{align}
	
	We assume that the microgrids are located close to each other, such that the loss of energy exchange is negligible. The new power balance constraint for each microgrid is written as
	\begin{equation}
		\begin{aligned}
			&& g_{i}^{t} + q_{b,i}^{t} + r_{d,i}^{t} + \sum_{j \in \mathcal{M} \backslash i} e_{i,j}^{t} = q_{s,i}^{t} + r_{c,i}^{t} + b_{i}^{t} + \sum_{n \in \mathcal{N}_i} x_{n}^{t},\\
			&& \forall t\in\mathcal{T}, \;\forall i \in\mathcal{M},\label{constraint-trading3}
		\end{aligned}
	\end{equation}
	where $\sum_{j \in \mathcal{M} \backslash i} e_{i,j}^{t}$ is the net energy traded between microgrid $i$ and all other microgrids in time slot $t$. If $\sum_{j \in \mathcal{M} \backslash i} e_{i,j}^{t} > 0$, microgrid $i$ purchases energy from other microgrids to serve its local demand; otherwise, $\sum_{j \in \mathcal{M} \backslash i} e_{i,j}^{t} < 0$, microgrid $i$ sells energy to make profit.
	
	Though the interconnected microgrids cooperate and trade energy with each other, they are still independent and selfish players. Each microgrid is a self-interested rational decision maker, it aims to optimize its own performance in terms of minimizing its total cost through energy trading. Compared with the cost function of microgrid $i$ in Problem \textbf{P-MG$_i$}, the interconnected microgrid has an extra cost that is the payment to other microgrids, \emph{i.e.}, 
	\begin{align*}
		C_{e} (\boldsymbol{\pi}_{i}) = \sum_{j \in \mathcal{M} \backslash i} \pi_{i,j}.
	\end{align*}
	
	The overall cost of microgrid $i$ consists of both operating cost \( \displaystyle C_i^{O}( \boldsymbol{q}_{i}, \boldsymbol{x}_{n}, \boldsymbol{r}_{c,i}, \boldsymbol{r}_{d,i}) \) and trading payment $C_{e} (\boldsymbol{\pi}_{i})$. It is clear that a microgrid will only participate in the trading if it can reduce its overall cost. This means that some microgrid may choose not participate in the trading. Thus, we have the following constraint:
	\begin{equation}
		\begin{aligned}
			C_i^{O}( \boldsymbol{q}_{i}, \boldsymbol{x}_{n}, \boldsymbol{r}_{c,i}, \boldsymbol{r}_{d,i}) + C_{e} (\boldsymbol{\pi}_{i})  \leq C_{i}^{Non}, ~ \forall i \in \mathcal{M}. \label{constraint-trading4}
		\end{aligned}
	\end{equation}
	
	Here the left-hand side of the inequality counts the overall cost of microgrid $i$ when participating in the energy trading. In the right-hand side, $C_{i}^{Non}$ denotes the minimized cost that microgrid $i$ can achieve without trading energy with other microgrids. In the bargaining literature we also call $C_{i}^{Non}$ the disagreement point.

	\subsection{Nash Bargaining Formulation for Energy Trading}
	We let $\mathcal{M}' \subseteq \mathcal{M}$ denote the set of microgrids who are willing to trade energy with each other. For the microgrids $j \in \mathcal{M} \backslash \mathcal{M}'$, they cannot benefit from energy trading and thus have no incentive to participate in the energy trading. Therefore, we focus on the microgrids in set $\mathcal{M}'$, and formulate their energy trading as a Nash bargaining problem:\\
	
	\textbf{Nash bargaining problem for energy trading (NBP)}
	\begin{align*}
		& \max && \prod_{i\in\mathcal{M}'} \left[ C_{i}^{Non} - \left( C_i^{O}( \boldsymbol{q}_{i}, \boldsymbol{x}_{n}, \boldsymbol{r}_{c,i}, \boldsymbol{r}_{d,i})  +  C_{e} (\boldsymbol{\pi}_{i}) \right) \right] \\
		& \text{subject to} 
		&& \text{\eqref{constraint-wind}--\eqref{constraint-sold}, \eqref{constraint-load1}, \eqref{constraint-load2}, \eqref{constraint-storage1}--\eqref{constraint-storage4}, and \eqref{constraint-selling}--\eqref{constraint-trading4}}, \\
		& \text{variables:} 
		&& \{ \boldsymbol{g}_{i}, \boldsymbol{q}_{i}, \boldsymbol{x}_{n}, \boldsymbol{r}_{c,i}, \boldsymbol{r}_{d,i}, \boldsymbol{e}_{i}, \boldsymbol{\pi}_{i}, ~i \in \mathcal{M}' \},
	\end{align*}
	where $C_{i}^{Non} - \left( C_i^{O}( \boldsymbol{q}_{i}, \boldsymbol{x}_{n}, \boldsymbol{r}_{c,i}, \boldsymbol{r}_{d,i})  +  C_{e} (\boldsymbol{\pi}_{i}) \right)$ corresponds to the cost reduction of microgrid $i$, \emph{i.e.}, the difference between disagreement point $C_{i}^{Non}$ and the total cost $C_i^{O}( \boldsymbol{q}_{i}, \boldsymbol{x}_{n}, \boldsymbol{r}_{c,i}, \boldsymbol{r}_{d,i})  +  C_{e} (\boldsymbol{\pi}_{i})$ of microgrid $i$ through bargaining. Compared with the summation of performance improvement of microgrids, the Nash product can guarantee that the benefits of cooperation are shared by each microgrid in a fair manner.
	
	Solving Problem \textbf{NBP} yields the optimal strategy of energy trading and payment, as well as the optimal power scheduling in each microgrid $i \in \mathcal{M}'$. However, we do not know the subset $\mathcal{M}'$ before hand. Moreover, to solve Problem \textbf{NBP} centrally, we need complete information of the interconnected-microgrids system, including all the operational parameters of each microgrid. This, however, may not be feasible in practice.
	
	To address the above two issues, we will analyze the Problem \textbf{NBP} to identify the microgrids with willingness to trade energy in Section IV, and design a decentralized solution method for the practical implementation in Section V.

	\section{Problem Analysis}
	In this Section, we will analyze the Nash bargaining problem \textbf{NBP} and explore the connection between the bargaining solution and the social optimal solution of the interconnected-microgrids system. Based on the connection, we then decompose \textbf{NBP} into two consecutive problems on energy trading \& scheduling and trading payment, respectively.

	We first assume that we know the subset $\mathcal{M}'$,\footnote{We will later discuss how to determine the subset $\mathcal{M}'$ in Theorem 1.} and focus on the microgrids in set $\mathcal{M}'$. Each microgrid $i \in \mathcal{M}'$ is willing to participate in the energy trading market and can strictly improve its performance in terms of lowering its total cost \emph{i.e.}, \( C_i^{O}( \boldsymbol{q}_{i}, \boldsymbol{x}_{n}, \boldsymbol{r}_{c,i}, \boldsymbol{r}_{d,i}) + C_{e} (\boldsymbol{\pi}_{i}) < C_{i}^{Non}  \). Take this group of microgrids as the new system, and the total cost for the system will be reduced due to energy trading, \emph{i.e.}, $ \sum_{i\in\mathcal{M}'} \left[ C_i^{O}( \boldsymbol{q}_{i}, \boldsymbol{x}_{n}, \boldsymbol{r}_{c,i}, \boldsymbol{r}_{d,i}) + C_{e} (\boldsymbol{\pi}_{i}) \right] < \sum_{i\in\mathcal{M}'} C_{i}^{Non} $. But there remains a question: what is the optimal total cost for the microgrids-system? We have the following observation presented in \textbf{Proposition} 1.
	
	\begin{proposition}
		For the microgrids in set $\mathcal{M}'$, the optimal energy trading \& scheduling solution to Problem \textbf{NBP} also minimizes the total cost of this group of microgrids $i \in \mathcal{M}'$.
	\end{proposition}
	
	\textbf{Proposition} 1 shows that the cooperation among microgrids achieves the best performance for the system. The intuition is as follows. The microgrids can be better off if the overall performance of the system improves, as they are able to gain benefits from the performance improvement (as cost reduction) of the overall system through proper money transfer (payments). Therefore, all the microgrids will cooperate to maximize the total benefit of the system, which explains why the minimum total cost can be achieved. For microgrids in set $\mathcal{M} \backslash \mathcal{M}'$, they do not have incentives to trade energy,\footnote{There are several scenarios where some microgrids do not have incentives to participate in the energy trading market. For example, when the microgrid perfectly balance its local demand and generation, there is no need to purchase energy from outside or no excessive supply to sell. Another possible scenario is that all the microgrids have excessive renewable energy generation than their demands, hence trading does not happen.} and thus each such microgrid $i$ achieves the same performance of $C_{i}^{Non}$ as in Section III.B, when they operate separately from other microgrids. Therefore, only those microgrids in set $\mathcal{M}'$ that trade energy with others contribute to the total cost reduction of the system. Next, we present a systematic method to determine which microgrids do not trade energy in the system.
	
	By \textbf{Proposition} 1, we can decompose the original Nash bargaining problem \textbf{NBP} into two sequential subproblems, as presented in \textbf{Theorem} 1.
	\begin{theorem}
		The energy-trading bargaining problem \textbf{NBP} can be decomposed into two sequential subproblems on energy trading \& scheduling and trading payment, denoted as \textbf{P1} and \textbf{P2}, respectively. In the first step, we solve a social operating cost minimization problem \textbf{P1} for the interconnected-microgrids system, which yields the optimal energy scheduling and energy trading for all microgrids. Those microgrids that trade energy with other microgrids form the set $\mathcal{M}'$. In the second step, microgrids in set $\mathcal{M}'$ bargain with each other to determine the energy-trading payments in \textbf{P2}.
	\end{theorem}
	
	\textbf{P1: Social operating cost minimization problem}
	\begin{equation*}
		\begin{aligned}
			& \min && \sum_{i\in\mathcal{M}} C_i^{O}( \boldsymbol{q}_{i}, \boldsymbol{x}_{n}, \boldsymbol{r}_{c,i}, \boldsymbol{r}_{d,i}) \\
			& \text{subject to} 
			&& \text{\eqref{constraint-wind}--\eqref{constraint-sold}, \eqref{constraint-load1}, \eqref{constraint-load2}, \eqref{constraint-storage1}--\eqref{constraint-storage4},  \eqref{constraint-selling}, \eqref{constraint-trading1}, and \eqref{constraint-trading3}},\\
			& \text{variables:}
			&&\{ \boldsymbol{g}_{i}, \boldsymbol{q}_{i}, \boldsymbol{x}_{n}, \boldsymbol{r}_{c,i}, \boldsymbol{r}_{d,i}, \boldsymbol{e}_{i}, ~i \in \mathcal{M} \}.
		\end{aligned} 
	\end{equation*}
	
	Solving \textbf{P1} determines the optimal energy trading \& scheduling $ \{\boldsymbol{g}_{i}^{\ast},\boldsymbol{q}_{i}^{\ast},\boldsymbol{x}_{n}^{\ast},\boldsymbol{r}_{c,i}^{\ast},\boldsymbol{r}_{d,i}^{\ast}, \boldsymbol{e}_{i}^{\ast} \}$ for each microgrid $i \in \mathcal{M}$. If a microgrid $i$ has a nonzero energy trading vector $\boldsymbol{e}_{i}$, then it belongs to set $\mathcal{M}'$, and will participate in the payment bargaining in Problem \textbf{P2}. For the non-trading microgrids $j \in \mathcal{M} \backslash \mathcal{M}'$, they do not participate in the payment bargaining and their operating costs are the same as the disagreement points $C_{j}^{Non}$.
	
	\textbf{P2: Payment bargaining problem}
	\begin{equation*}
		\begin{aligned}
			& \max 
			&& \prod_{i\in\mathcal{M}'} \Big( \delta_{i}^{\ast} -  C_{e} (\boldsymbol{\pi}_{i}) \Big)  \\
			& \text{subject to} 
			&& \text{\eqref{constraint-trading2} and \eqref{constraint-trading4}},\\
			& \text{variables:} 
			&& \{ \boldsymbol{\pi}_{i},~i \in \mathcal{M}' \},
		\end{aligned}
	\end{equation*}
	where \( \displaystyle \delta_{i}^{\ast} \triangleq C_{i}^{Non} - C_i^{O}( \boldsymbol{q}_{i}^{\ast}, \boldsymbol{x}_{n}^{\ast}, \boldsymbol{r}_{c,i}^{\ast}, \boldsymbol{r}_{d,i}^{\ast}) \) denotes the operating cost reduction of microgrid $i$ based on the optimal solution of Problem \textbf{P1}.
	
	\textbf{Theorem} 1 decomposes the bargaining problem \textbf{NBP} into two subproblems, which can be solved sequentially. Firstly, the interconnected microgrids cooperate together to trade energy and minimize the social operating cost of the entire system. Any microgrid trading energy with others contributes to the social cost reduction of the entire multi-microgrid system. These microgrids can benefit from energy trading through sharing the reduced social cost via the bargaining, and hence have incentives to participate in the trading. Specifically, microgrids selling energy to others can receive payments through bargaining, and microgrids buying energy from others can reduce their operational costs (though they need to pay for energy purchase). Solving Problem \textbf{P1} involves all the microgrids and does not exclude any of microgrids in the system. In the optimal solution of Problem \textbf{P1}, if there exist any microgrids who do not trade with others, these non-trading microgrids do not contribute to the social cost minimization and won't benefit from energy trading. The only criterion for determining trading and non-trading microgrids is whether they can contribute to the social cost minimization. Secondly, after identifying those microgrids who do not want to participate in the energy trading, we then focus on the rest of the microgrids who trade energy. We decide a fair allocation of benefits to each of participated microgrids through the payment bargaining problem \textbf{P2}.
	
	Problem \textbf{P1} and \textbf{P2} are both convex, and can be solved efficiently in a centralized optimization manner. However, this may not be feasible in practice, as each microgrid is an independent decision maker, and the smart grid operator may not directly control each microgrid's energy trading and scheduling. Moreover, each microgrid has two categories of decision variables: power schedules \( \displaystyle  \{ \boldsymbol{g}_{i}, \boldsymbol{q}_{i}, \boldsymbol{x}_{n}, \boldsymbol{r}_{c,i}, \boldsymbol{r}_{d,i} \} \) as internal variables, and energy trading \& payment $\{ \boldsymbol{e}_{i}, \boldsymbol{\pi}_{i} \}$ as external variables. Centralized optimization is not practical, because it may violate the privacy of microgrid's internal operation. We will discuss the design of a decentralized algorithm to solve \textbf{P1} and \textbf{P2} in the next section.
	
	For the proofs of Theorems, please refer to the appendix.
	
	\section{Decentralized Solution Method}
	In this section, we design a decentralized solution method, which enables the microgrids to coordinate with each other to solve \textbf{P1} and \textbf{P2}. We use the alternating direction method of multipliers (ADMM) \cite{ADMM} to design the distributed algorithm, as ADMM has good convergence properties for the optimization problems with non-strictly convex objective functions and large-scale variables. 
	
	\subsection{Solving P1 (Social Cost Minimization)}
	First, we solve \textbf{P1} in a decentralized fashion. Since the convergence of the multi-block ADMM algorithm cannot be guaranteed, we introduce auxiliary variables for the energy trading decisions in order to convert the $M$-block (corresponding to $M$-microgrid) structure of the original optimization problem into an equivalent two-block structure. The convergence of ADMM algorithm with two blocks of variables is guaranteed \cite{ADMM}. Specifically, let us introduce auxiliary variables $\boldsymbol{\hat{e}}_{i} = \{\hat{e}_{i,j}^{t},\forall j \in \mathcal{M} \backslash i, \forall t \in \mathcal{T}\}$, and replace constraints \eqref{constraint-trading1} by
	\begin{align}
		& \hat{e}_{i,j}^{t} = e_{i,j}^{t},~\forall t \in \mathcal{T},~\forall j \in \mathcal{M} \backslash i,~\forall i \in \mathcal{M}, \label{constraint-auxiliary1}\\
		& \hat{e}_{i,j}^{t} + \hat{e}_{j,i}^{t} = 0,~\forall t \in \mathcal{T},~\forall j \in \mathcal{M} \backslash i,~\forall i \in \mathcal{M}. \label{constraint-auxiliary2}
	\end{align}
	
	We define $\boldsymbol{\lambda} = \{ \lambda_{i,j}^{t} \}$ as the dual variables associated with constraints \eqref{constraint-auxiliary1}, and have the augmented Lagrangian for \textbf{P1}:
	\begin{align*}
		& \mathcal{L}_1 ( \boldsymbol{g}_{i}, \boldsymbol{q}_{i}, \boldsymbol{x}_{n}, \boldsymbol{r}_{c,i}, \boldsymbol{r}_{d,i}, \boldsymbol{e}_{i}, \boldsymbol{\hat{e}}_{i}, \boldsymbol{\lambda} ) \\
		= & \sum_{i\in\mathcal{M}} 
		\Big[ 
		C_i^{O}( \boldsymbol{q}_{i}, \boldsymbol{x}_{n}, \boldsymbol{r}_{c,i}, \boldsymbol{r}_{d,i}) \\
		+ & \sum_{j \in \mathcal{M} \backslash i} \sum_{t\in\mathcal{T}} 
		\left( \frac{\rho_{1}}{2} \left( \hat{e}_{i,j}^{t} - e_{i,j}^{t} \right)^{2} 
		+ \lambda_{i,j}^{t} \left( \hat{e}_{i,j}^{t} - e_{i,j}^{t} \right) 
		\right) 
		\Big],
	\end{align*}
	where $\rho_{1} >0$ is a parameter for the quadratic penalty of constraint \eqref{constraint-auxiliary1}.

	The ADMM solution method involves iterations between a lower level problem and a higher level problem. Specifically, the lower level problem involves microgrids solving their local optimization problems in parallel based on fixed dual variables $\boldsymbol{\lambda}$ and auxiliary variables $\boldsymbol{\hat{e}}$. The upper level problem involves updating the auxiliary variables and dual variables using the results from the low level problems. 
	
	In iteration $k$, given parameter $\rho_{1}(k)$, dual variables $\lambda_{i,j}^{t}(k)$ and auxiliary variables $\hat{e}_{i,j}^{t}(k)$, each microgrid $i$ solves its local optimization problem:
	
	\textbf{Local optimization problem of P1 (P1-MG$_i$)}
	\begin{equation*}
		\begin{aligned}
			& \min 
			&& C_i^{O}( \boldsymbol{q}_{i}, \boldsymbol{x}_{n}, \boldsymbol{r}_{c,i}, \boldsymbol{r}_{d,i}) \\
			&&& + \sum_{j \in \mathcal{M} \backslash i} \sum_{t\in\mathcal{T}} 
			\left( \frac{\rho_{1}}{2} \left( \hat{e}_{i,j}^{t}(k) - e_{i,j}^{t} \right)^{2} - \lambda_{i,j}^{t}(k) e_{i,j}^{t}  \right)\\
			& \text{subject to}
			&&  \text{\eqref{constraint-wind}--\eqref{constraint-sold}, \eqref{constraint-load1}, \eqref{constraint-load2}, \eqref{constraint-storage1}--\eqref{constraint-storage4}, \eqref{constraint-selling}, and \eqref{constraint-trading3}},\\
			& \text{variables:}
			&& \boldsymbol{g}_{i}, \boldsymbol{q}_{i}, \boldsymbol{x}_{n}, \boldsymbol{r}_{c,i}, \boldsymbol{r}_{d,i}, \boldsymbol{e}_{i}.
		\end{aligned} 
	\end{equation*}

	Based on the solution $\boldsymbol{e}_{i}(k+1)$ from Problem \textbf{P1-MG$_i$}, the higher level problem updates auxiliary variables $\boldsymbol{\hat{e}}_{i}$ and dual variables $\boldsymbol{\lambda}$. Specifically, the higher level problem of \textbf{P1} is formulated as follows:
	\begin{equation*}
		\begin{aligned}
			& \min  && \sum_{i \in \mathcal{M}} \sum_{j \in \mathcal{M} \backslash i} \sum_{t\in\mathcal{T}} 
			\Big(
			\frac{\rho_{1}}{2} \left( \hat{e}_{i,j}^{t} - e_{i,j}^{t}(k+1) \right)^{2} \\
			&&& + \lambda_{i,j}^{t}(k) \hat{e}_{i,j}^{t} 
			\Big)\\
			& \text{subject to} 
			&& \text{\eqref{constraint-auxiliary2}}, \\
			& \text{variables:} 
			&& \{ \boldsymbol{\hat{e}}_{i},~i \in \mathcal{M} \}.
		\end{aligned} 
	\end{equation*}
	
	As the auxiliary variables are only coupled between each pair of trading partners, we can solve the higher level problem of \textbf{P1} by considering the following optimization problem only involving a pair of microgrids $i$ and $j$,
		\begin{equation*}
			\begin{aligned}
				& \min  
				&& \frac{\rho_{1}}{2} \left( \hat{e}_{i,j}^{t} - e_{i,j}^{t}(k+1) \right)^{2} + \lambda_{i,j}^{t}(k) \hat{e}_{i,j}^{t} \\
				&&&  + 
				\frac{\rho_{1}}{2} \left( \hat{e}_{j,i}^{t} - e_{j,i}^{t}(k+1) \right)^{2} + \lambda_{j,i}^{t}(k) \hat{e}_{j,i}^{t} \\
				& \text{subject to} 
				&& \hat{e}_{i,j}^{t} + \hat{e}_{j,i}^{t} = 0, \\
				& \text{variables:} 
				&& \{ \hat{e}_{i,j}^{t}, \hat{e}_{j,i}^{t} \},
			\end{aligned}
		\end{equation*}
		and obtain the optimal closed-form solution for updating $\hat{e}_{i,j}^{t}$:
		\begin{align}
			\begin{split}
				& \hat{e}_{i,j}^{t}(k+1) = -\hat{e}_{j,i}^{t}(k+1) \\
				= & \frac{\rho_{1} \left( e_{i,j}^{t}(k+1) - e_{j,i}^{t}(k+1) \right) - \left( \lambda_{i,j}^{t}(k) - \lambda_{j,i}^{t}(k) \right) }{2 \rho_{1}}. \label{updateenergy}
			\end{split}
		\end{align}
	
	Based on $e_{i,j}^{t}(k+1)$ and $\hat{e}_{i,j}^{t}(k+1)$, we update the dual variables as follows:
	\begin{align}
		\lambda_{i,j}^{t}(k+1) = \lambda_{i,j}^{t}(k) + \rho_{1} \left( \hat{e}_{i,j}^{t}(k+1) - e_{i,j}^{t}(k+1) \right) . \label{updatelambda}
	\end{align}

	\subsection{Solving P2 (Payment Bargaining)}
	Second, we solve \textbf{P2} in a decentralized fashion. We introduce auxiliary variables $\boldsymbol{\hat{\pi}}_{i} = \{\hat{\pi}_{i,j},\forall j \in \mathcal{M} \backslash i\}$, and replace constraints \eqref{constraint-trading2} by
	\begin{align}
		& \hat{\pi}_{i,j} = \pi_{i,j},~\forall j \in \mathcal{M} \backslash i,~\forall i \in \mathcal{M}, \label{constraint-auxiliary3}\\
		& \hat{\pi}_{i,j} + \hat{\pi}_{j,i} = 0,~\forall j \in \mathcal{M} \backslash i,~\forall i \in \mathcal{M}. \label{constraint-auxiliary4}
	\end{align}
	
	We take the log transformation of the objective function of the bargaining problem \textbf{P2}, and write the Lagrangian for \textbf{P2}:
	\begin{align*}
		\mathcal{L}_{2} (\boldsymbol{\pi}_{i},\boldsymbol{\hat{\pi}}_{i},\boldsymbol{\gamma}) 
		& = \sum_{i\in\mathcal{M}} \Big[ 
		- \ln \Big( \delta_{i}^{\ast} - C_{e} (\boldsymbol{\pi}_{i}) \Big) \\
		&  + \sum_{j \in \mathcal{M} \backslash i} 
		\left(   \frac{\rho_{2}}{2} \left( \hat{\pi}_{i,j} - \pi_{i,j} \right)^{2} 
		+ \gamma_{i,j}  \left( \hat{\pi}_{i,j} - \pi_{i,j} \right)
		\right) 
		\Big],
	\end{align*}
	where $\boldsymbol{\gamma} = \{ \gamma_{i,j}\}$ are dual variables associated with constraints \eqref{constraint-auxiliary3}, and $\rho_{2}$ is a penalty parameter.

	Similarly, solving Problem \textbf{P2} involves iterations between a lower problem and a higher problem. In iteration $k$, given $\hat{\pi}_{i,j}(k)$ and $\gamma_{i,j}(k)$, each microgrid solves its local optimization problem:
	
	\textbf{Local optimization problem of P2 (P2-MG$_i$)}
	\begin{equation*}
		\begin{aligned}
			& \min
			&& - \ln \Big( 
			\delta_{i}^{\ast} - C_{e} (\boldsymbol{\pi}_{i}) \Big) \\
			&&& + \sum_{j \in \mathcal{M} \backslash i}  
			\left(   
			\frac{\rho_{2}}{2} \left( \hat{\pi}_{i,j}(k) - \pi_{i,j}  \right)^{2} 
			- \gamma_{i,j}(k)  \pi_{i,j}
			\right)  \\
			& \text{subject to} 
			&& \text{\eqref{constraint-trading4}},\\
			& \text{variables:} 
			&& \boldsymbol{\pi}_{i}.
		\end{aligned}
	\end{equation*}
	
	Based on the solution $\pi_{i,j}(k+1)$ from Problem \textbf{P2-MG$_i$}, the higher level problem updates the auxiliary variables $\boldsymbol{\hat{\pi}}_{i}$ and dual variable $\boldsymbol{\gamma}$. Specifically, the higher level problem of \textbf{P2} is formulated as follows:
	\begin{equation*}
		\begin{aligned}
			& \min  && \sum_{i \in \mathcal{M}} \sum_{j \in \mathcal{M} \backslash i} 
			\left(
			\frac{\rho_{2}}{2} \left( \hat{\pi}_{i,j} - \pi_{i,j}(k+1) \right)^{2} 
			+ \gamma_{i,j}(k)  \hat{\pi}_{i,j} \right)\\
			& \text{subject to} 
			&& \eqref{constraint-auxiliary4},\\
			& \text{variables:} 
			&& \{ \boldsymbol{\hat{\pi}}_{i},~i \in \mathcal{M} \}.
		\end{aligned}
	\end{equation*}
	
	We can solve the higher level problem of \textbf{P2} by considering a pair of microgrids $i$ and $j$,
		\begin{equation*}
			\begin{aligned}
				& \min  
				&& \frac{\rho_{2}}{2} \left( \hat{\pi}_{i,j} - \pi_{i,j}(k+1) \right)^{2} 
				+ \gamma_{i,j}(k)  \hat{\pi}_{i,j} \\
				&&& +
				\frac{\rho_{2}}{2} \left( \hat{\pi}_{j,i} - \pi_{j,i}(k+1) \right)^{2} 
				+ \gamma_{j,i}(k)  \hat{\pi}_{j,i}  \\
				& \text{subject to} 
				&& \hat{\pi}_{i,j} + \hat{\pi}_{j,i} = 0, \\
				& \text{variables:} 
				&& \{ \hat{\pi}_{i,j}, \hat{\pi}_{j,i} \},
			\end{aligned}
		\end{equation*}
		and obtain the optimal closed-form solution for updating $\hat{\pi}_{i,j}$:
		\begin{align}
			\begin{split}
				& \hat{\pi}_{i,j}(k+1) = -\hat{\pi}_{j,i}(k+1) \\
				= & \frac{\rho_{2} \left( \pi_{i,j}(k+1) - \pi_{j,i}(k+1) \right) - \left( \gamma_{i,j}(k) - \gamma_{j,i}(k) \right) }{2 \rho_{2}}. \label{updatepayment}
			\end{split}
		\end{align}

	Based on $\pi_{i,j}(k+1)$ and $\hat{\pi}_{i,j}(k+1)$, we update the dual variables as follows:
	\begin{align}
		\gamma_{i,j}(k+1) = \gamma_{i,j}(k) + \rho_{2} 
		\left( \hat{\pi}_{i,j}(k+1) - \pi_{i,j}(k+1) \right) . \label{updategamma}
	\end{align}

	\subsection{Algorithm Design and Implementation}
	
	To implement the solution method for solving the energy trading and payment problems \textbf{P1} and \textbf{P2}, we design a distributed algorithm, as shown in \textbf{Algorithm 1}. We see that \textbf{Algorithm 1} solves problems \textbf{P1} and \textbf{P2} through iterations between higher level and lower level problems within Step 1 and Step 2, respectively. We introduce the concept of virtual clearing house for solving the higher level problem. The virtual clearing house is a communication and computing module\footnote{The virtual clearing house is a non-profit driven computing module. The virtual clearing house can communicate with all participating microgrids using smart grid communication technologies, \textit{e.g.}, cellular-based wide area networks and power line communications. The virtual clearing house is able to protect the privacy of all the microgrids.}, which provides clearing service of energy trading and payment for all the microgrids. At the beginning of each daily operation, the virtual clearing house updates the auxiliary and dual variables according to (\ref{updateenergy}), (\ref{updatelambda}) and (\ref{updatepayment}), (\ref{updategamma}), and broadcasts to all participating microgrids when solving \textbf{P1} and \textbf{P2}, respectively. All the participating microgrids concurrently solve \textbf{P1-MG$_i$} and \textbf{P2-MG$_i$} and report their energy trading schedules and payments to the virtual clearing house. 
	
	We see that \textbf{Algorithm 1} only requires limited information exchange overhead among microgrids. Each microgrid only needs to communicate with the clearing house instead of all the remaining microgrids. Such communication can be supported by many existing one-to-many communication technologies, such as the LTE cellular technology. For the microgrids, they only need to report their energy trading and payment schedules, as it's reasonable to let the counterpart microgrids know the trading amount of energy and payment during each iteration. As for the local power scheduling, demand response, and energy storage charging/discharging, each microgrid will compute its own optimal decision, without disclosing its private information of internal operations to other microgrids. Therefore, \textbf{Algorithm 1} solves the energy trading problem \textbf{EP1} with minimum information without releasing microgrid's private power schedules.

	\begin{algorithm}[!htb] 
		\caption{Distributed algorithm solving \textbf{P1} and \textbf{P2}}
		\label{alg1}
		\begin{algorithmic}[1]
			\State \textbf{Step 1}: solve Problem \textbf{P1}.
			\State ~~Initialization: iteration index $k=1$, error tolerance $\epsilon_{1} > 0$, stepsize $\rho_{1}(0) =1$, and initial multipliers $\boldsymbol{\lambda}(0) =\textbf{0}$.
			\Repeat
			\State At $k$-th iteration,
			\State \textbf{Lower Level Problem:} Microgrid $i$ solves Problem \textbf{P1-MG$_i$} based on the current value of dual variables $\boldsymbol{\lambda}(k)$;
			\State \textbf{Higher Level Problem:} The virtual clearing house makes update according to (\ref{updateenergy}) and (\ref{updatelambda}).
			\State Update iteration index $k=k+1$;
			\Until terminal condition is satisfied, \emph{i.e.}, $\sum_{i \in \mathcal{M}} \parallel \boldsymbol{\hat{e}}_{i}(k) - \boldsymbol{e}_{i}(k) \parallel \leq \epsilon_{1}. $
			\State \textbf{Step 2}: solve Problem \textbf{P2}. 
			\State Initialization: iteration index $k=1$, error tolerance $\epsilon_{2} > 0$, stepsize $\rho_{2}(0) =1$, and initial multipliers $\boldsymbol{\gamma}(0) = \textbf{0}$.
			\Repeat
			\State At $k$-th iteration,
			\State \textbf{Lower Level Problem:} Microgrid $i$ solves Problem \textbf{P2-MG$_i$} based on the current value of dual variables $\boldsymbol{\gamma}(k)$;
			\State \textbf{Higher Level Problem:} The virtual clearing house makes updates according to the rules in (\ref{updatepayment}) and (\ref{updategamma}).
			\State Update iteration index $k=k+1$;
			\Until terminal condition is satisfied, \emph{i.e.}, $\sum_{i \in \mathcal{M}'} \parallel \boldsymbol{\hat{\pi}}_{i}(k) - \boldsymbol{\pi}_{i}(k) \parallel \leq \epsilon_{2}. $
			\State \textbf{end}
		\end{algorithmic}
	\end{algorithm}

	Based on \cite{ADMM}, we demonstrate the convergence of \textbf{Algorithm 1} in the following theorem.
	\begin{theorem}
		\textbf{Algorithm 1} converges to the optimal solution of \textbf{P1} in step 1 and the optimal solution of \textbf{P2} in step 2, under proper stepsizes of $\rho_{1}(k)$ and $\rho_{2}(k)  $, e.g., $\rho_{1}(k) = 1/k \to 0$, and $\rho_{2}(k) = 1/k \to 0$, as $k \to \infty$.
	\end{theorem}
	
	We can divide the decision variables of Problem \textbf{P1} into two blocks: $ \{ \boldsymbol{g}_{i}, \boldsymbol{q}_{i}, \boldsymbol{x}_{n}, \boldsymbol{r}_{c,i}, \boldsymbol{r}_{d,i}, \boldsymbol{e}_{i}, ~i \in \mathcal{M} \} $ and $ \{ \boldsymbol{\hat{e}}_{i}, ~i \in \mathcal{M} \} $. Similarly, we can divide the decision variables of Problem \textbf{P2} into two blocks: $ \{\boldsymbol{\pi}_{i}, ~i \in \mathcal{M}' \} $ and $ \{\boldsymbol{\hat{\pi}}_{i}, ~i \in \mathcal{M}' \} $. Since \textbf{P1} and \textbf{P2} are convex problems and \textbf{Algorithm 1} is a standard two-block ADMM algorithm, we conclude that the convergence of \textbf{Algorithm 1} is guaranteed \cite{ADMM}.

	\section{Simulation Evaluations}
\begin{figure*} 
		\centering
		\begin{minipage}[t]{0.45\linewidth}
			\centering
			\includegraphics[width=5.5cm]{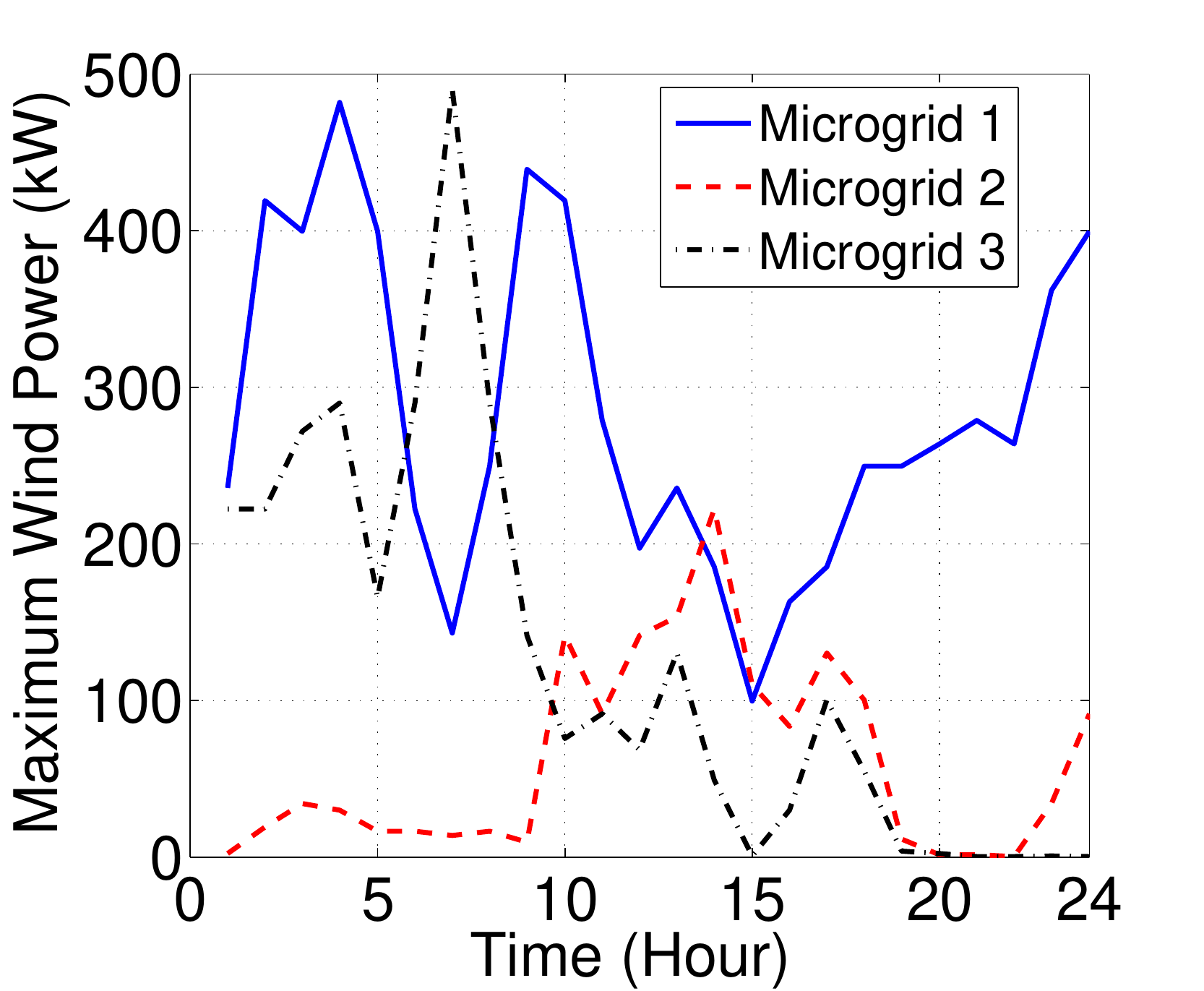}
			\caption{\label{fig-windpower} Wind power in Microgrid 1, 2 and 3.}
		\end{minipage}
		\begin{minipage}[t]{0.05\linewidth}
			\centering
		\end{minipage}
		\begin{minipage}[t]{0.45\linewidth}
			\centering
			\includegraphics[width=5.5cm]{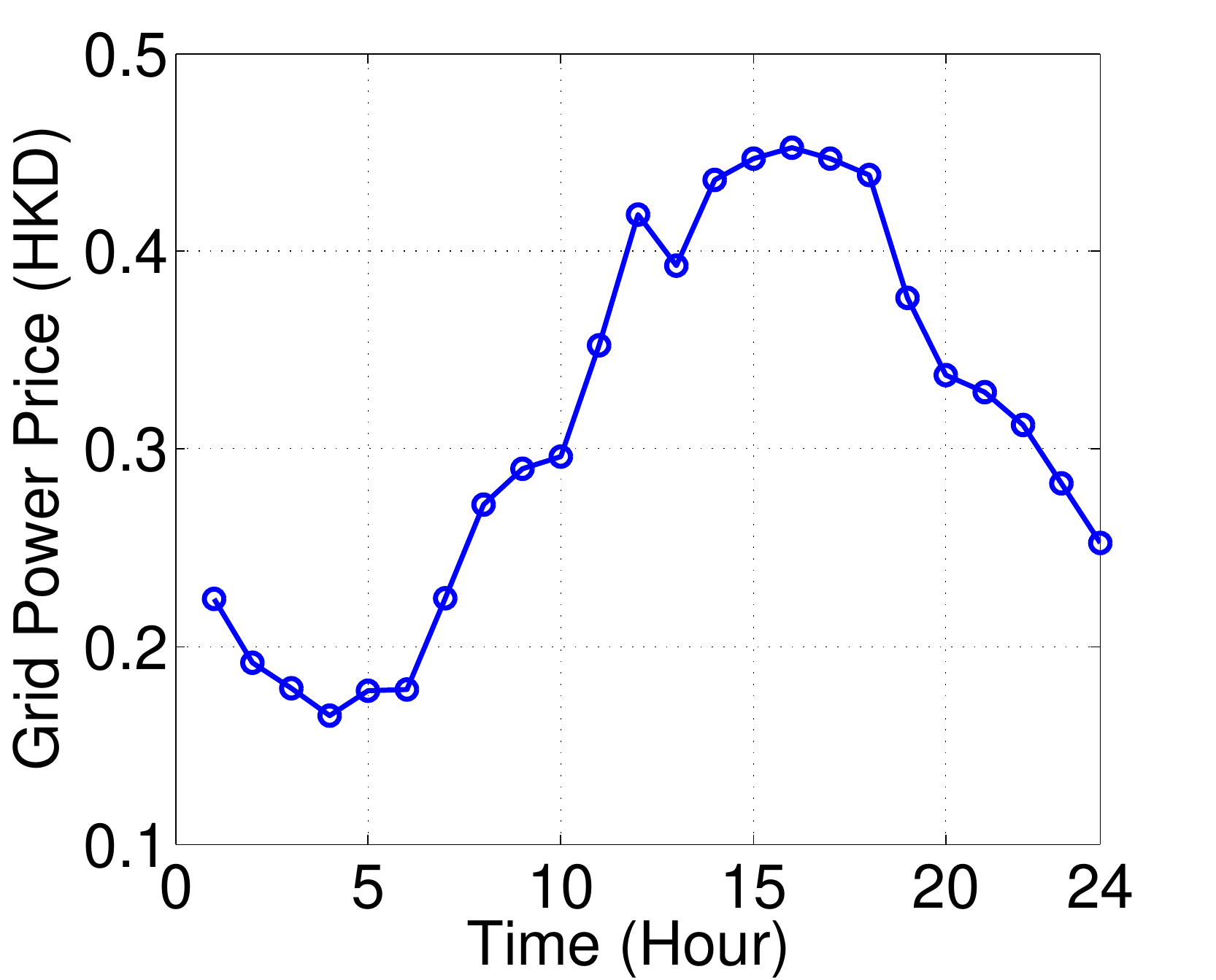}
			\caption{\label{fig-price} Power grid price.}
		\end{minipage}
	\end{figure*}

	We consider three interconnected microgrids, each having its local wind generation. Based on the hourly wind speed data \cite{hkdata} at several different Hong Kong locations, we calculate the daily realizations of hourly wind power, and use wind power productions on January 18, 2013 at Tate's Cairn, Tai Po Kau, and Sai Kung of Hong Kong as local renewable generations in the three microgrids, respectively. Fig. \ref{fig-windpower} depicts the wind power generations in microgrids 1, 2 and 3. The electricity price of the main power grid is retrieved from ISO New England \cite{price} and is depicted in Fig. \ref{fig-price}, and the feed-in rate is set as 0.1. Other parameters are summarized as follows: $G_{1} = 600$, $G_{2} = G_{3} = 1000$, $Q_{1}^{ \max} = 500$, $Q_{2}^{ \max} = Q_{2}^{ \max} = 300$, $\beta_{1} = 1.0 $, $\beta_{1} = \beta_{2} = 0.5$, $r_{c,1}^{ \max} = r_{d,1}^{ \max} = 30$, $r_{c,2}^{ \max} = r_{d,2}^{ \max} = 40$, $r_{c,3}^{ \max} = r_{d,3}^{ \max} = 50$, $S_{1}^{ \max}=100$, $S_{2}^{ \max} = S_{3}^{ \max} = 200$, $c_{s} = 0.01$, and $\eta_{c,i} = \eta_{d,i} = 0.95$, $i=1,2,3$.

	\subsection{Optimal Energy Trading}
\begin{figure*}
		\centering
		\begin{subfigure}[b]{0.3\textwidth}
			\includegraphics[width=\textwidth]{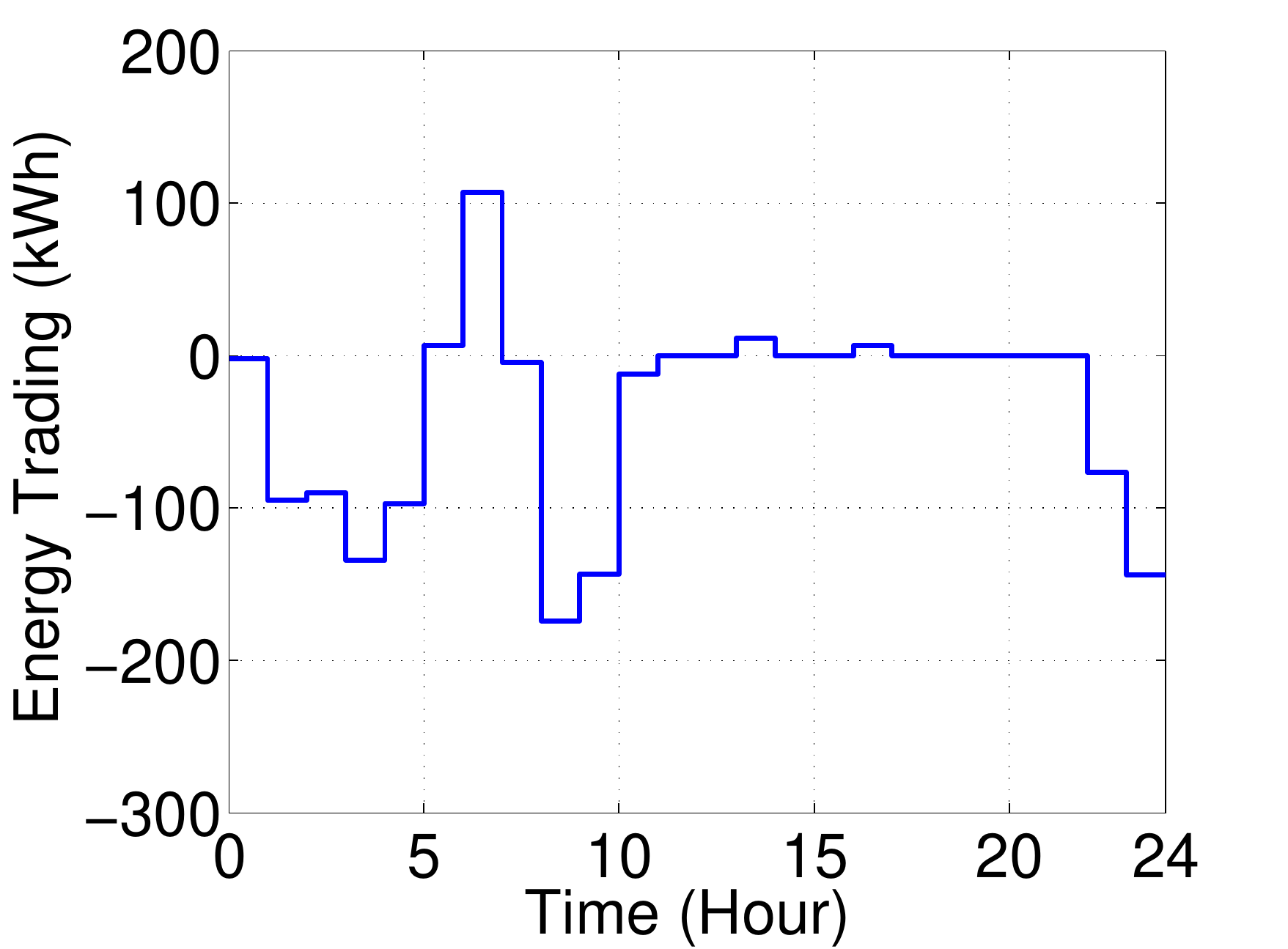}
			\caption{\label{fig-trading1} Microgrid 1}
		\end{subfigure}
		~ 
		\begin{subfigure}[b]{0.3\textwidth}
			\includegraphics[width=\textwidth]{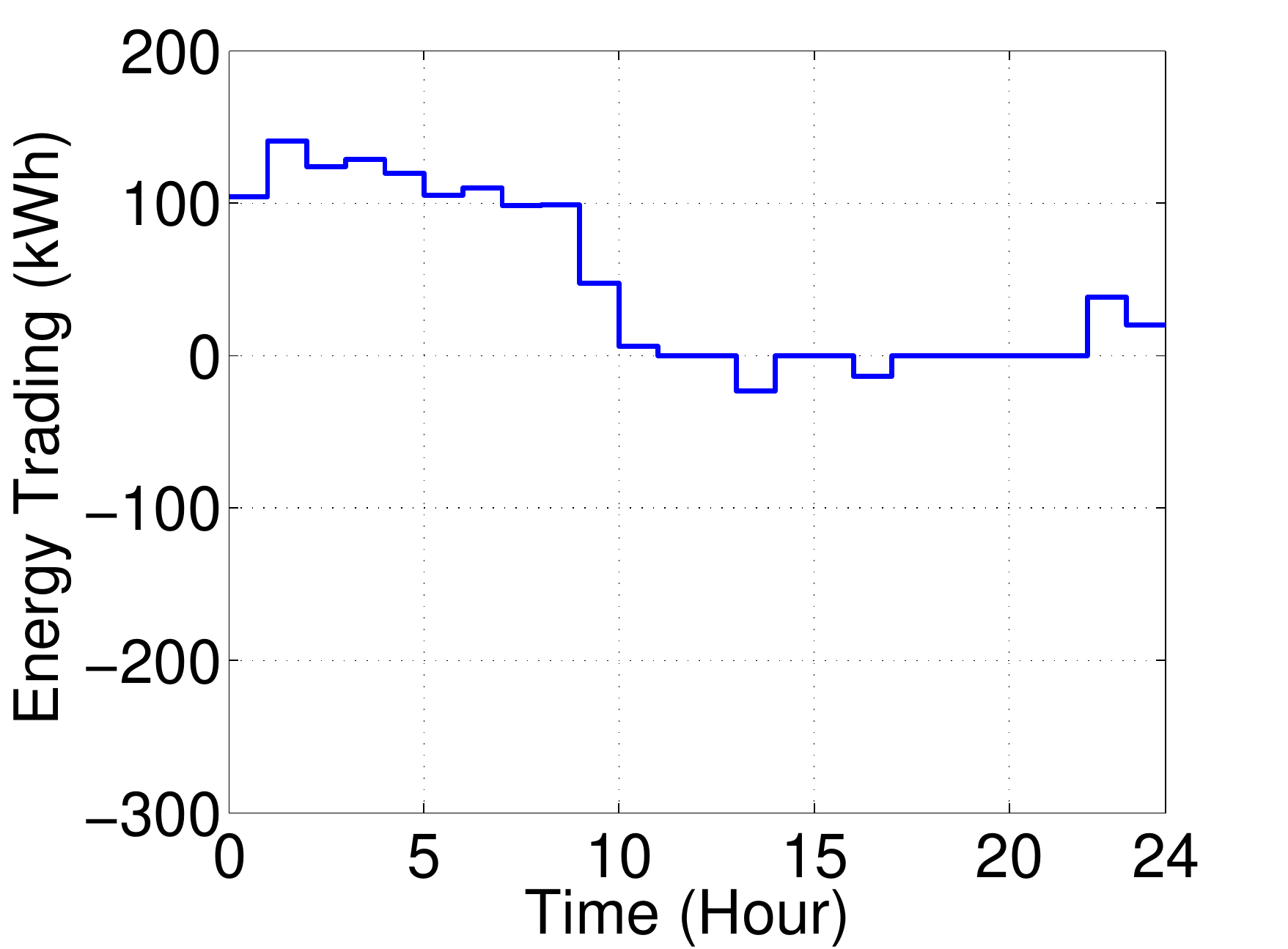}
			\caption{\label{fig-trading2} Microgrid 2}
		\end{subfigure}
		~ 
		\begin{subfigure}[b]{0.3\textwidth}
			\includegraphics[width=\textwidth]{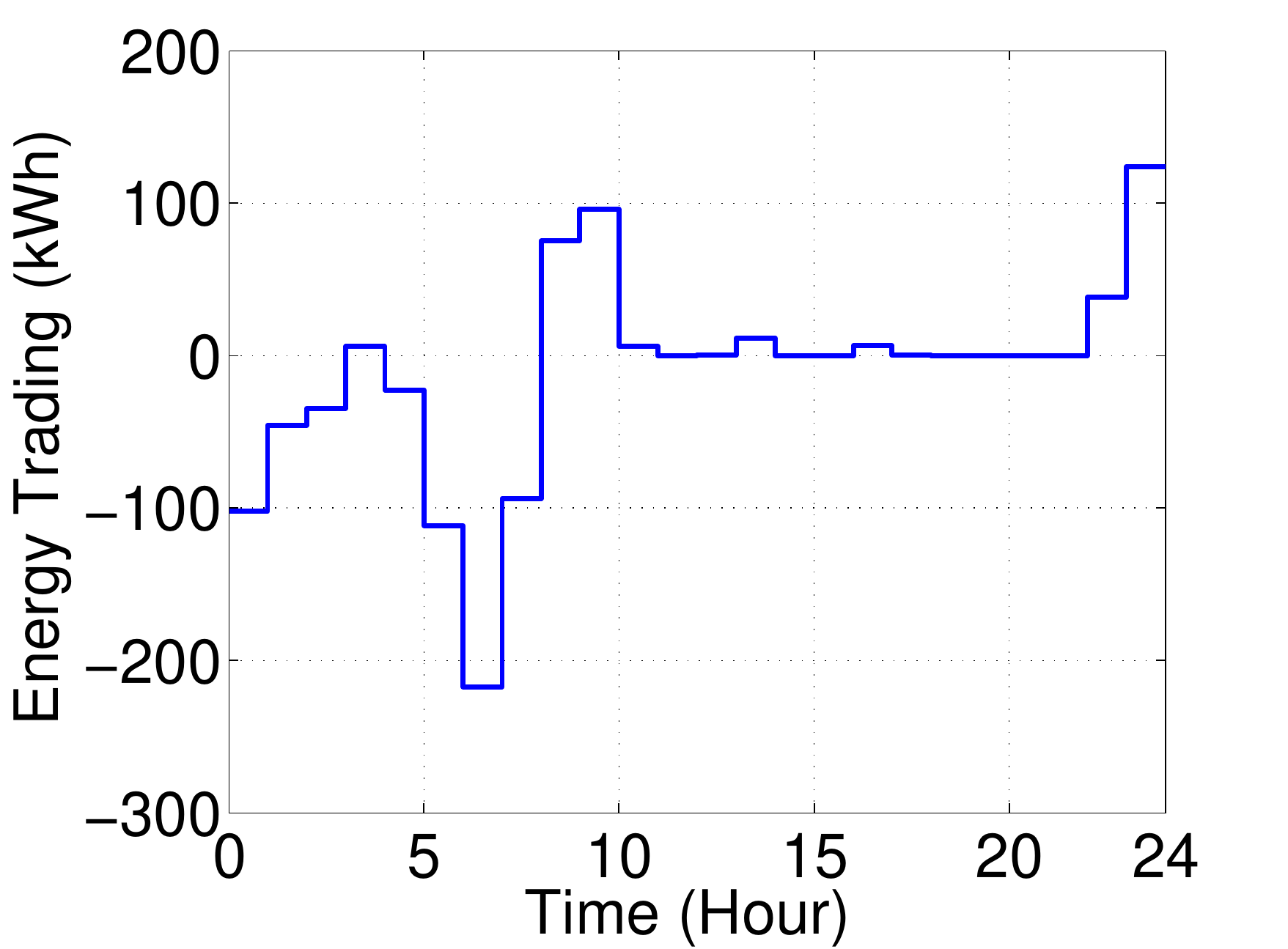}
			\caption{\label{fig-trading3} Microgrid 3}
		\end{subfigure}
		\caption{\label{fig-trading} Energy trading of Microgrid 1, 2 and 3.}
	\end{figure*}

	We first study the optimal energy trading of all three microgrids, as depicted in Fig. \ref{fig-trading}. Here positive values correspond to purchasing energy, and negative values correspond to selling energy. We see that all three microgrids exchange energy actively across the 24-hour operation horizon. From Fig. \ref{fig-windpower}, we see that microgrid 1 has higher wind power output than other microgrids, and thus sell excessive energy to other microgrids during most of time slots except hour 7. The reason for this is that microgrid 1 has a sudden drop of wind power supply in hour 7, while microgrid 3 has adequate generation in the same hour. Therefore, microgrid 1 purchases energy from microgrid 3 in hour 7. Microgrid 2 and microgrid 3 purchase energy during hours 1-10 and 23-24, because they lack local wind generations during the corresponding time slots. 

	\subsection{Optimal Power Scheduling}
\begin{figure*}
		\centering
		\begin{subfigure}[b]{0.3\textwidth}
			\includegraphics[width=\textwidth]{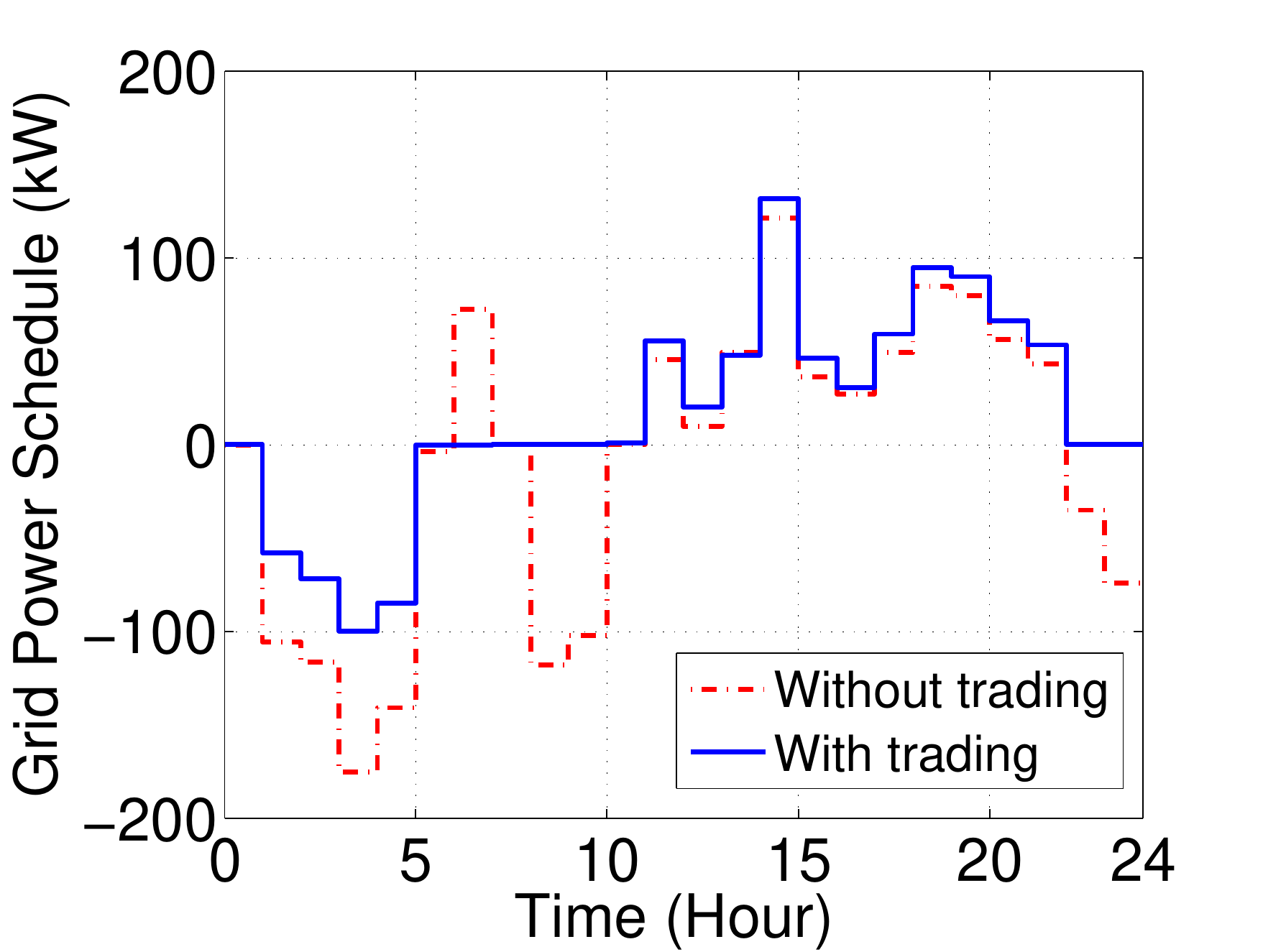}
			\caption{\label{fig-gridpower1} Microgrid 1}
		\end{subfigure}
		~ 
		\begin{subfigure}[b]{0.3\textwidth}
			\includegraphics[width=\textwidth]{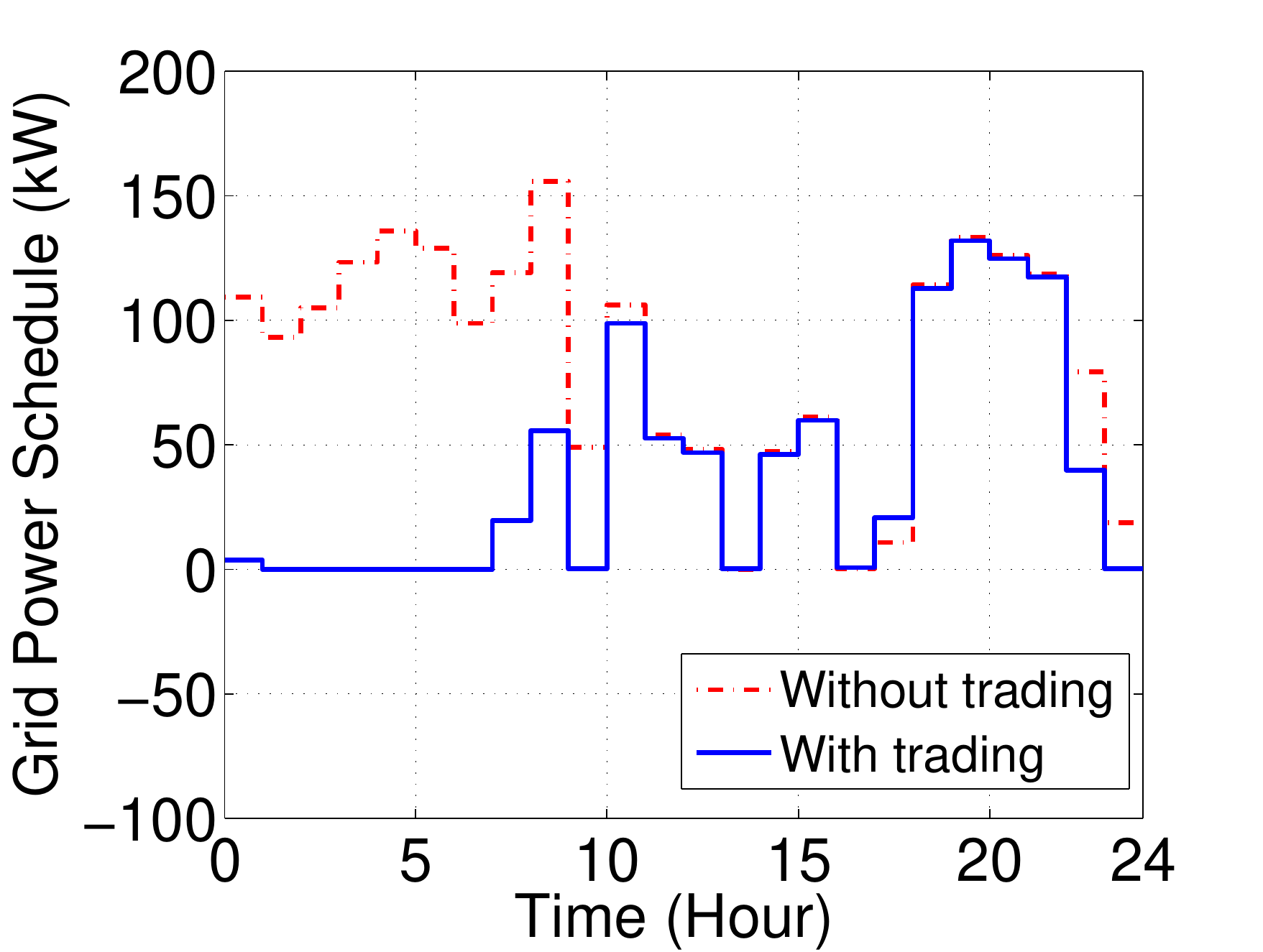}
			\caption{\label{fig-gridpower2} Microgrid 2}
		\end{subfigure}
		~ 
		\begin{subfigure}[b]{0.3\textwidth}
			\includegraphics[width=\textwidth]{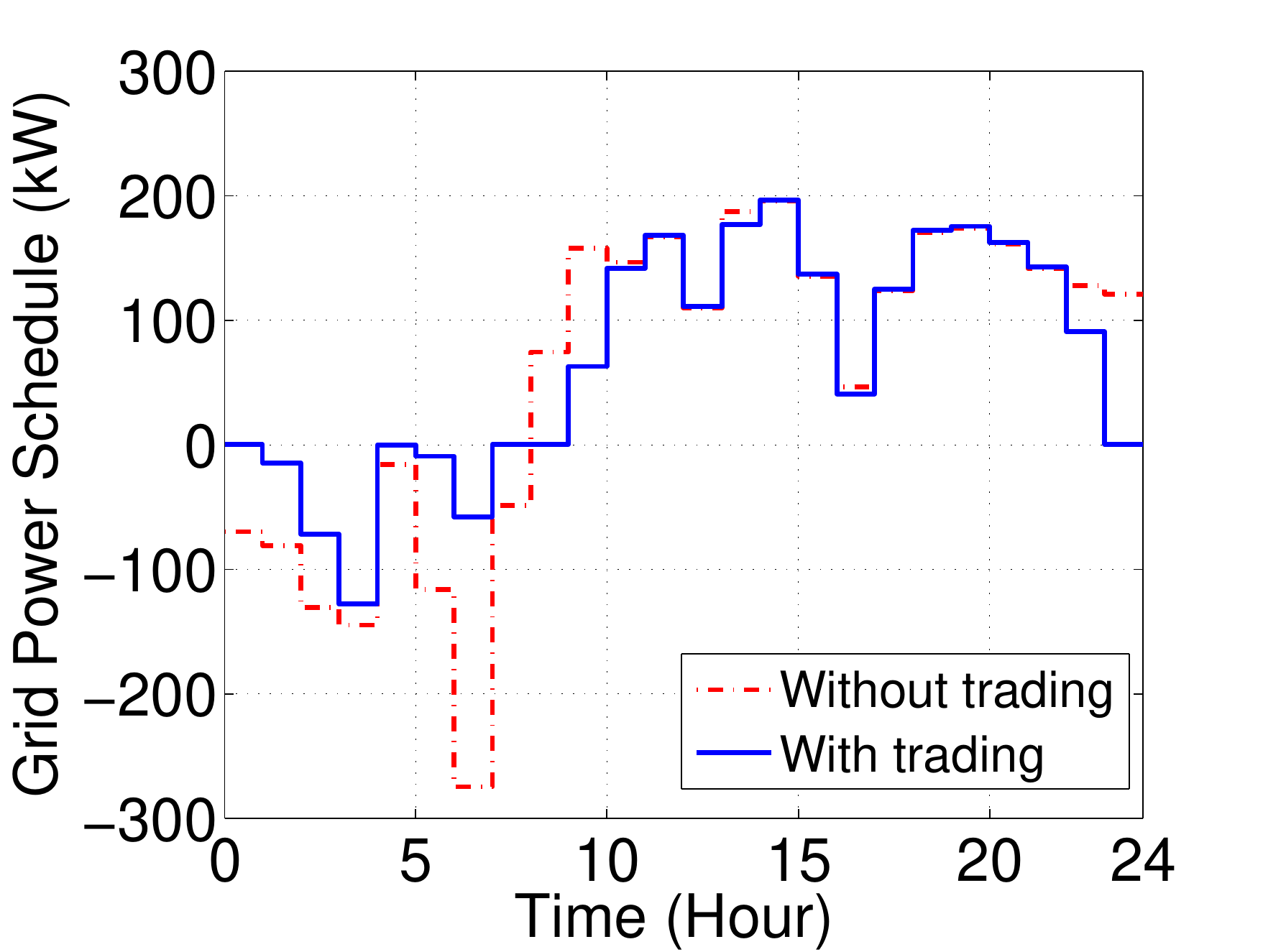}
			\caption{\label{fig-gridpower3} Microgrid 3}
		\end{subfigure}
		\caption{\label{fig-gridpower} Gird power procurement of Microgrid 1, 2 and 3.}
	\end{figure*}

	Through energy trading, the overall multi-microgrid system purchases less power from the main grid, hence improves the overall utilization of local renewable energy. We compare the main grid power schedules with and without energy trading, as depicted in Fig. \ref{fig-gridpower}. Note that even without energy trading among microgrids, we always allow microgrids to sell excessive renewable energy to the main grid. We see that microgrid 2 purchases less power from main grid with energy trading, because they can get power supply from microgrid 1 and microgrid 3. Microgrid 1 and microgrid 3 also sell less power to the main grid with energy trading, because they sell more local wind power to microgrid 2.

\begin{figure*}
		\centering
		\begin{subfigure}[b]{0.3\textwidth}
			\includegraphics[width=\textwidth]{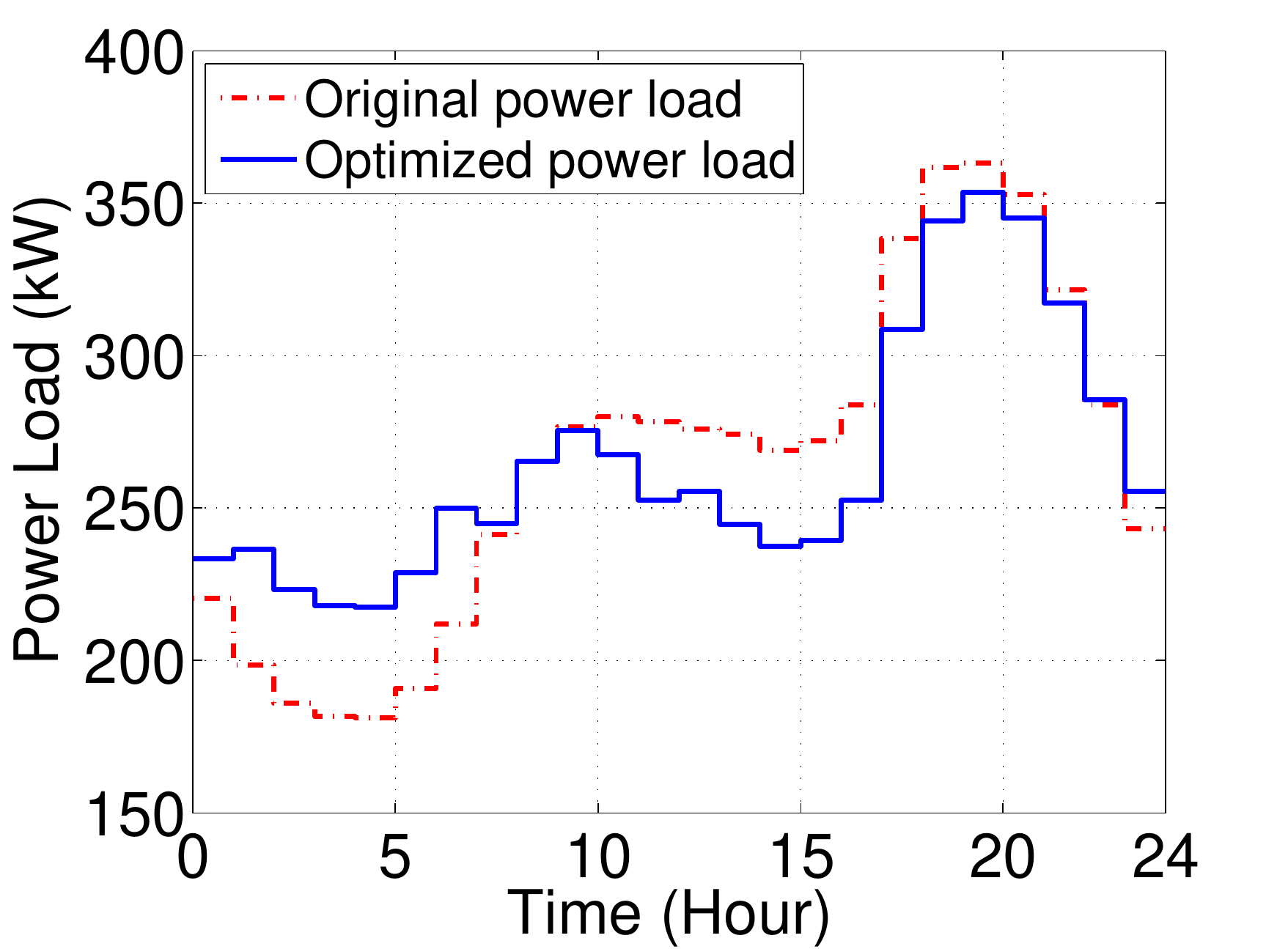}
			\caption{\label{fig-powerload1} Microgrid 1}
		\end{subfigure}
		~ 
		\begin{subfigure}[b]{0.3\textwidth}
			\includegraphics[width=\textwidth]{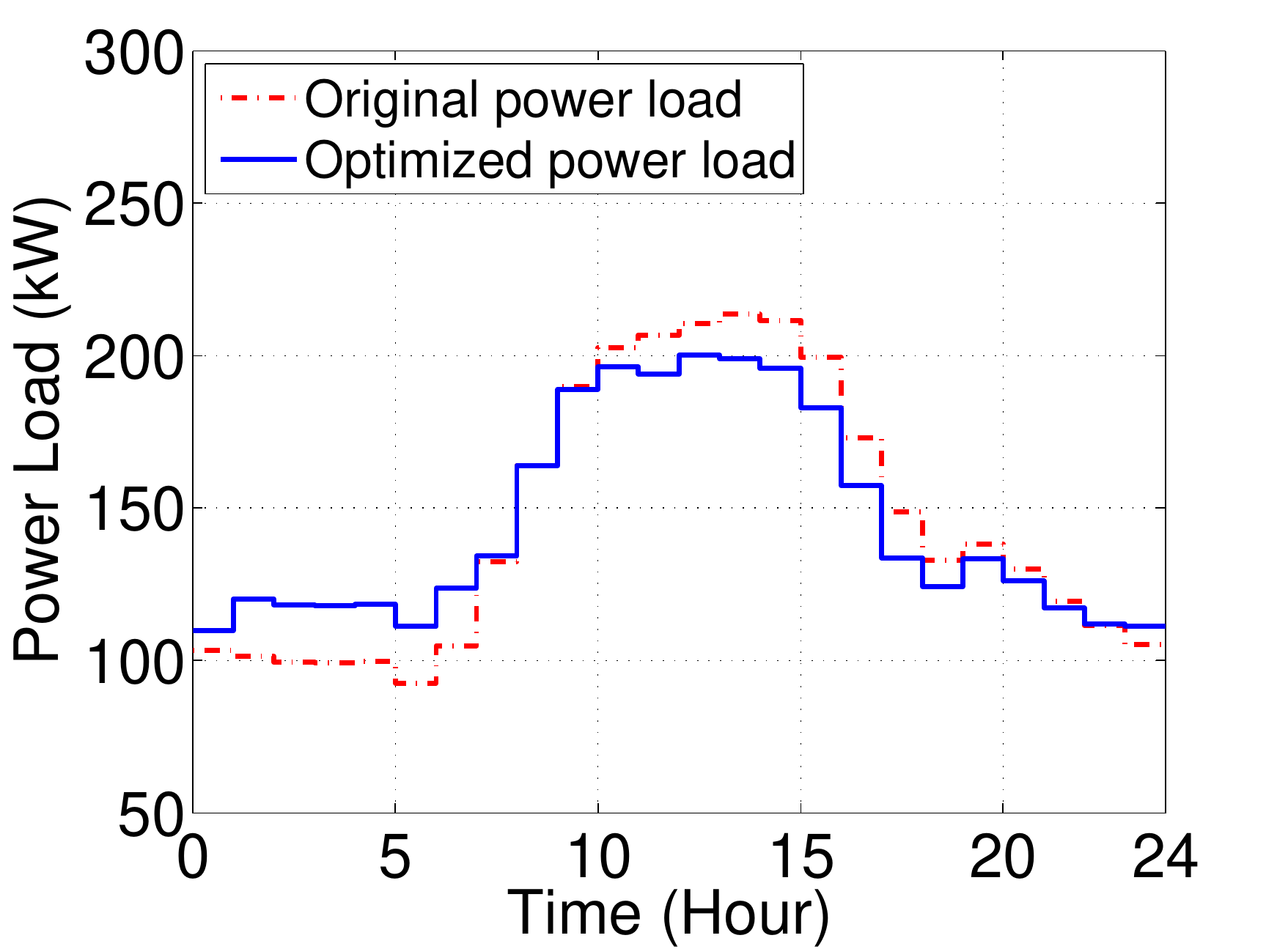}
			\caption{\label{fig-powerload2} Microgrid 2}
		\end{subfigure}
		~ 
		\begin{subfigure}[b]{0.3\textwidth}
			\includegraphics[width=\textwidth]{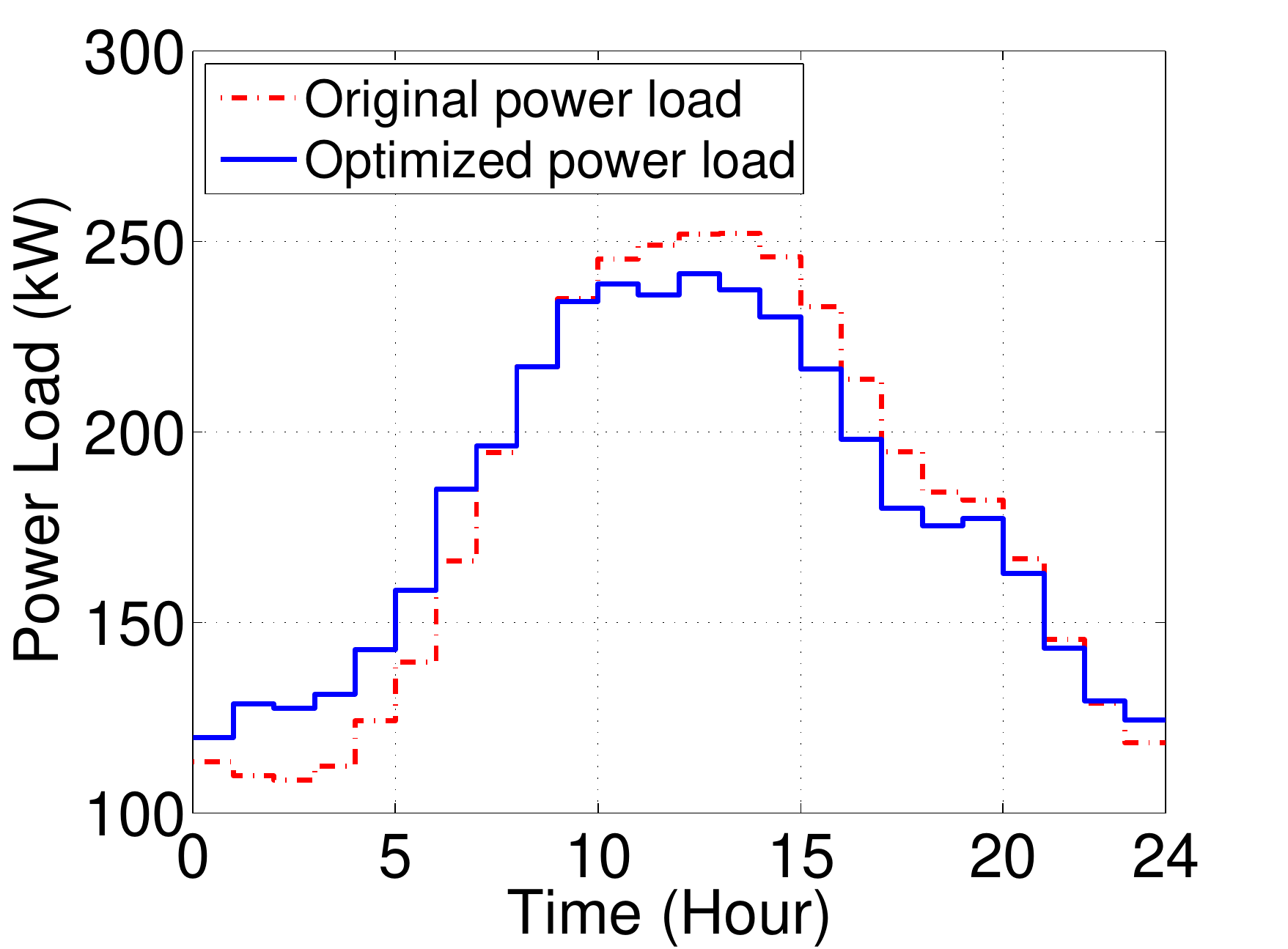}
			\caption{\label{fig-powerload3} Microgrid 3}
		\end{subfigure}
		\caption{\label{fig-powerload} Demand response in Microgrid 1, 2 and 3.}
	\end{figure*}

	Fig. \ref{fig-powerload} depicts the optimal demand response of microgrids 1, 2 and 3 in the scenario with energy trading. We see that microgrid 1 has peak power load at night time (hours 18-22), and original power loads in microgrid 2 and microgrid 3 achieve peak levels at day time (hours 9-17). As shown in Fig. \ref{fig-windpower} and Fig. \ref{fig-price}, the aggregate renewable power generation has higher output during hours 1-10, and the electricity price is higher in peak-time slots (hours 11-20). Therefore, all the microgrids shift their flexible loads from peak-time slots to off-peak time slots to use more renewable energy and reduce their operational costs.
	
\begin{figure*}
		\centering
		\begin{subfigure}[b]{0.3\textwidth}
			\includegraphics[width=\textwidth]{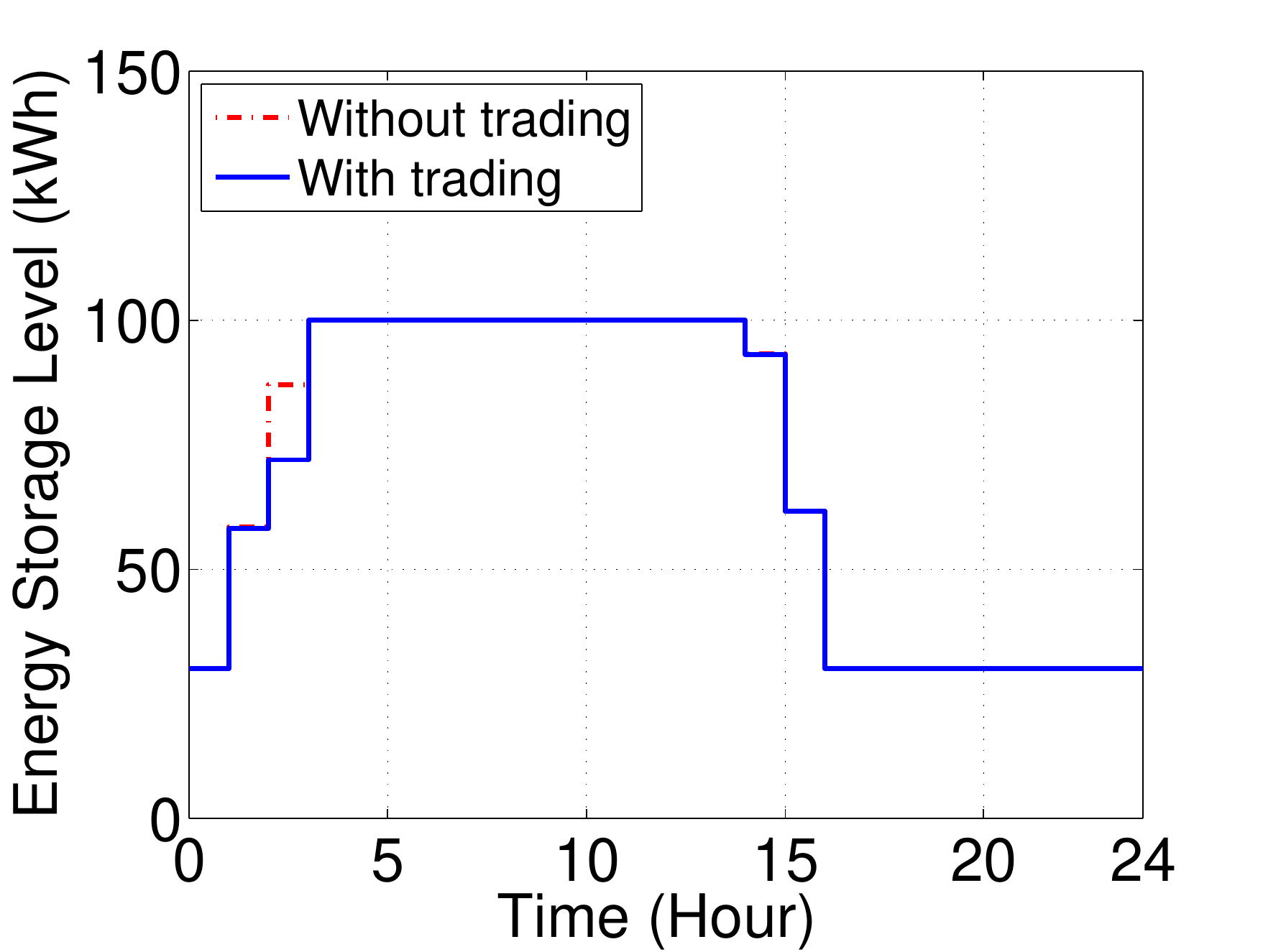}
			\caption{\label{fig-storage1} Microgrid 1}
		\end{subfigure}
		~ 
		\begin{subfigure}[b]{0.3\textwidth}
			\includegraphics[width=\textwidth]{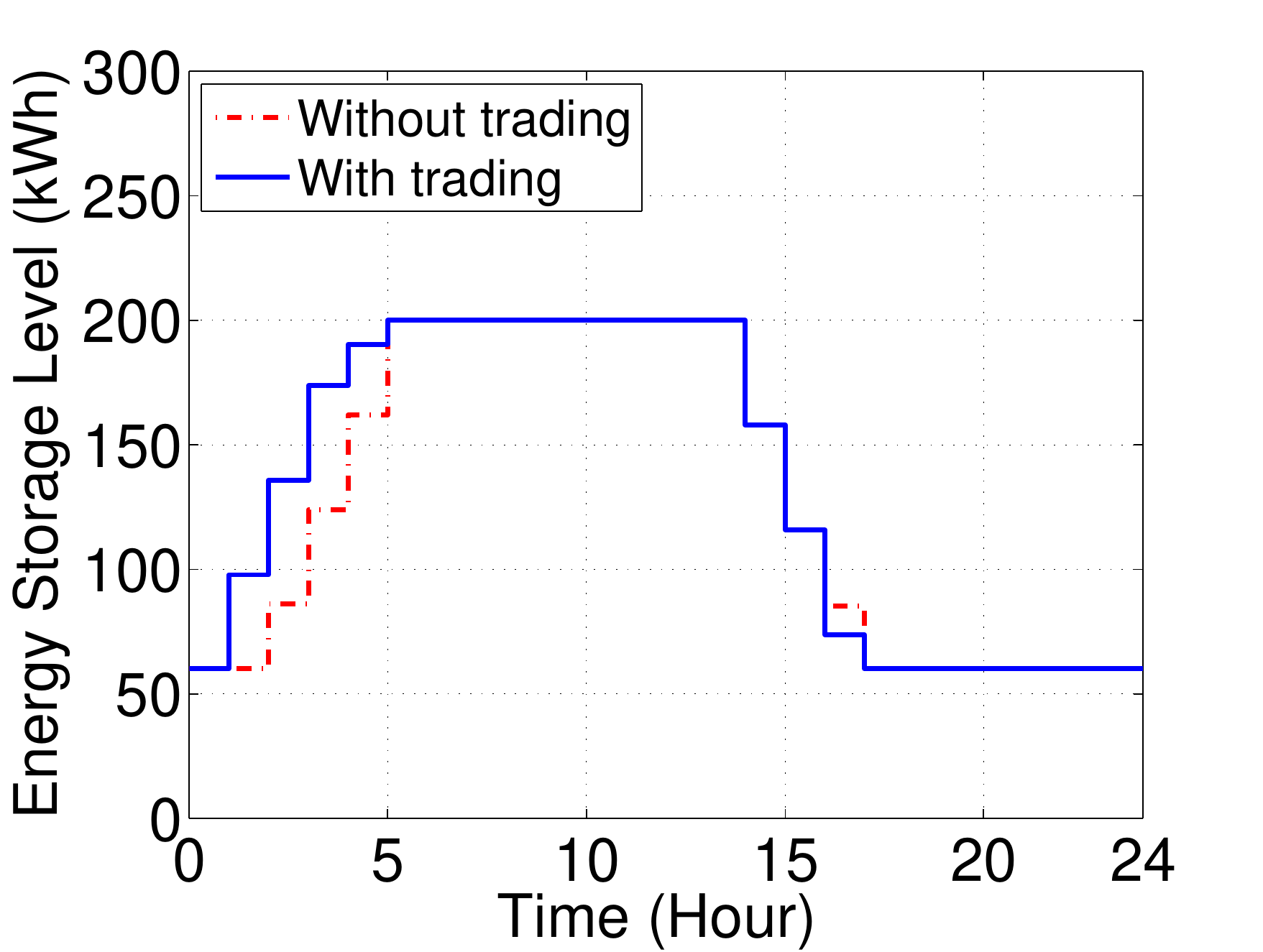}
			\caption{\label{fig-storage2} Microgrid 2}
		\end{subfigure}
		~ 
		\begin{subfigure}[b]{0.3\textwidth}
			\includegraphics[width=\textwidth]{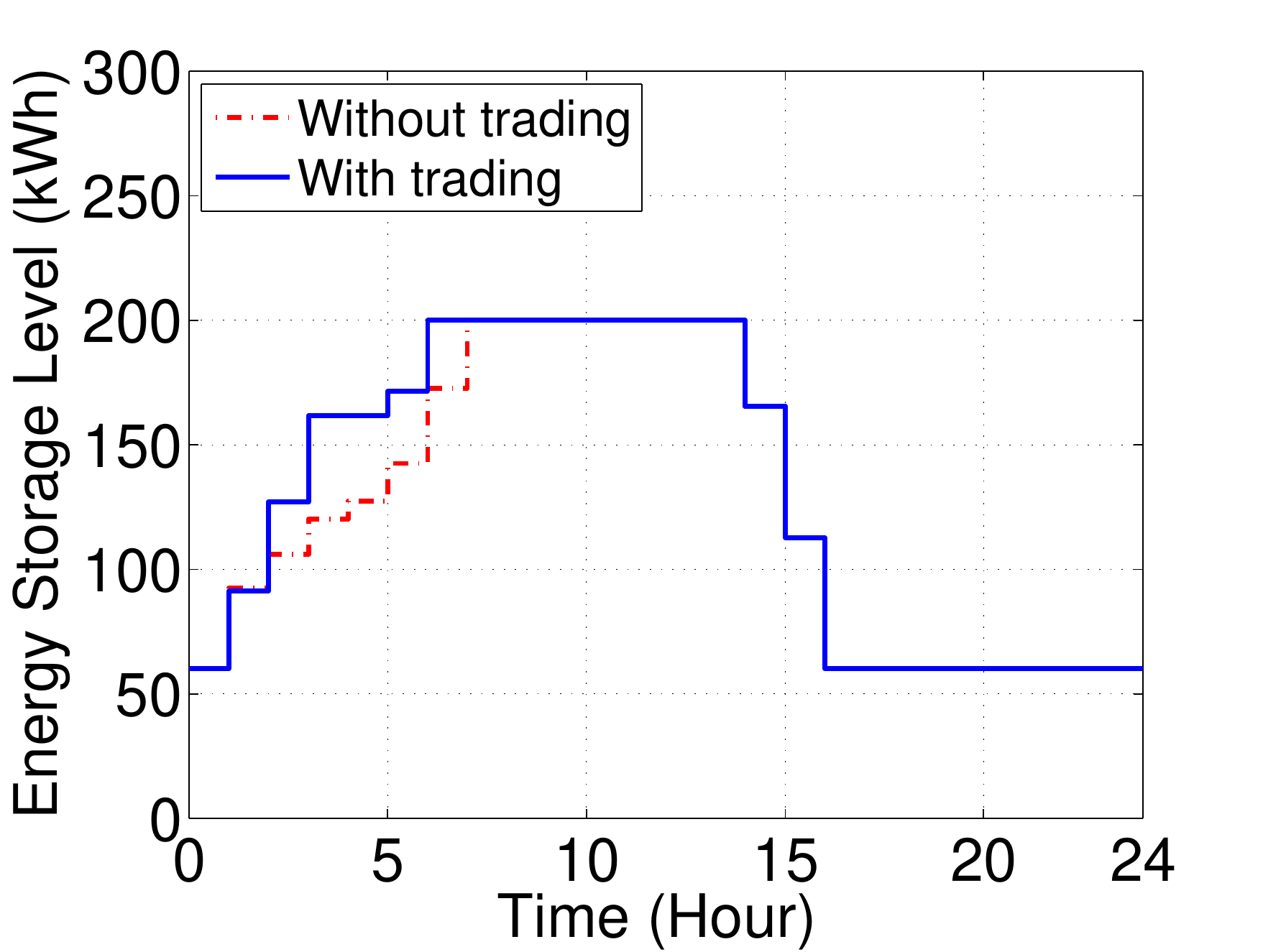}
			\caption{\label{fig-storage3} Microgrid 3}
		\end{subfigure}
		\caption{\label{fig-storage} Energy storage dynamics in Microgrid 1, 2 and 3.}
	\end{figure*}

	Fig. \ref{fig-storage} shows the comparison of optimal energy storage dynamics with and without energy trading, for microgrids 1, 2 and 3, respectively. We see that at the beginning of the day, all microgrids charge more power into energy storage in the scenario with energy trading than without energy trading. This is because microgrids can trade renewable power with other microgrids, and thus store more energy in batteries to meet their peak loads later.
	
	\subsection{Optimal Payment}
	Table \ref{tab-payment} shows the operating costs of microgrids 1, 2 and 3 and their payments for energy trading. The total cost of the interconnected microgrids is reduced by up to 13.2\% from 1637.8 to 1422.4. We see that energy trading reduces the operating costs of microgrids 2 and 3. The operating cost of microgrid 1, however, increases. The reason is that microgrid 1 sells more wind power to other microgrids for profits, instead of selling back to the main grid. Microgrid 2 pays 124.5 and 33.4 to microgrid 1 and microgrid 3. Considering the overall impact of cost and payment, every microgrid benefits through energy trading. For example, microgrid 1 reduces the cost from 243.8 (no trading) to cost plus payment 172.1 (with trading), which corresponds to a 29.4\% decrease. This demonstrates the effectiveness of our proposed payment scheme, which incentivizes microgrids to participate in the energy trading.
	
	\begin{table}[!htb] 
		\caption{Costs and payments}
		\centering
		\begin{tabular}{c| c| c| c| c}
			\hline \hline
			Merits vs Microgrids &  1 &  2 &  3 & System \\
			\hline 	 	 
			Cost (no trading) & 243.8 & 607.0 & 787.0 & 1637.8  \\ 	 	
			Cost (with trading) & 296.5  & 377.4 & 748.6  & 1422.4 	 \\ 	 	
			Payment (for trading) & -124.5  & 157.8 & -33.4 & 0.0 \\
			Cost + Payment (with trading) & 172.1  & 535.2  & 715.2  & 1422.4 \\
			\hline
		\end{tabular}
		\label{tab-payment}
	\end{table}

	\section{Conclusion}
	In this paper, we studied the energy trading problem among interconnected microgrids using Nash bargaining theory. We first presented a model for a single microgrid that captures key features of smart grid technologies. Then we designed a bargaining-based energy trading market for interconnected microgrids, and proposed a distributed solution method. Numerical simulations based on realistic meteorological data demonstrated the effectiveness of the energy trading mechanism. In our future work, we will consider the role of the power grid operator in the energy trading of microgrids.

	\appendix
	\subsection{Proof of Proposition 1}
	For the microgrids in set $\mathcal{M}'$, they have incentives to trade energy with each other. For any group of microgrids who trade energy with each other, we conclude that the aggregated operational cost of the microgrids-group $\mathcal{G}$ decreases, \emph{i.e.}, \( \sum_{i \in \mathcal{G}} \left( C_i^{O}( \boldsymbol{q}_{i}, \boldsymbol{x}_{n}, \boldsymbol{r}_{c,i}, \boldsymbol{r}_{d,i}) - C_{i}^{Non} \right) <0 \). There always exists a feasible payment allocation $ \{ \boldsymbol{\pi}_{i}, i \in \mathcal{G} \}$, such that its performance improves in terms of cost reduction, \emph{i.e.}, \( C_i^{O}( \boldsymbol{q}_{i}, \boldsymbol{x}_{n}, \boldsymbol{r}_{c,i}, \boldsymbol{r}_{d,i}) + C_{e} (\boldsymbol{\pi}_{i}) < C_{i}^{Non}  \).
	
	As constraint \eqref{constraint-trading4} is guaranteed to be satisfied, we can divide the decision variables of the Nash bargaining problem \textbf{NBP} into two sets: the energy trading \& power scheduling variables $ \{ \boldsymbol{g}_{i}, \boldsymbol{q}_{i}, \boldsymbol{x}_{n}, \boldsymbol{r}_{c,i}, \boldsymbol{r}_{d,i}, \boldsymbol{e}_{i}, ~i \in \mathcal{M}' \} $ and trading payment variables $ \{  \boldsymbol{\pi}_{i}, ~i \in \mathcal{M}' \} $. We observe that these two sets of variables are decoupled, thus we can solve Problem \textbf{NBP} in a sequential manner.
	
	Given the optimal energy trading \& power scheduling decisions $ \{ \boldsymbol{g}_{i}^{\ast}, \boldsymbol{q}_{i}^{\ast}, \boldsymbol{x}_{n}^{\ast}, \boldsymbol{r}_{c,i}^{\ast}, \boldsymbol{r}_{d,i}^{\ast}, \boldsymbol{e}_{i}^{\ast}, ~i \in \mathcal{M}' \} $, we can solve the optimal trading payment decisions $ \{  \boldsymbol{\pi}_{i}^{\ast}, ~i \in \mathcal{M}' \} $ through
		\begin{align*}
			& \max && \prod_{i\in\mathcal{M}'} \left[ C_{i}^{Non} - \left( C_i^{O}( \boldsymbol{q}_{i}^{\ast}, \boldsymbol{x}_{n}^{\ast}, \boldsymbol{r}_{c,i}^{\ast}, \boldsymbol{r}_{d,i}^{\ast})  +  
			C_{e} (\boldsymbol{\pi}_{i}) \right) \right] \\
			& \text{subject to} 
			&& \text{\eqref{constraint-trading2}},\\
			& \text{variables:} 
			&& \{ \boldsymbol{\pi}_{i}, ~i \in \mathcal{M}' \},
		\end{align*}
		and obtain 
		\begin{equation}
			\begin{split}
				C_{e} (\boldsymbol{\pi}_{i}^{\ast} ) = & C_{i}^{Non} - C_i^{O}( \boldsymbol{q}_{i}^{\ast}, \boldsymbol{x}_{n}^{\ast}, \boldsymbol{r}_{c,i}^{\ast}, \boldsymbol{r}_{d,i}^{\ast}) \\
				- & \frac{ \sum_{i\in\mathcal{M}'} \left( C_{i}^{Non} -  C_i^{O}( \boldsymbol{q}_{i}^{\ast}, \boldsymbol{x}_{n}^{\ast}, \boldsymbol{r}_{c,i}^{\ast}, \boldsymbol{r}_{d,i}^{\ast}) \right) } {M'}. \label{optimalpayment} 
			\end{split}
		\end{equation}
	
	Substituting (\ref{optimalpayment}) into Problem \textbf{NBP} yields the optimal objective:
		\begin{equation}
			\left[  
			\frac{\sum_{i\in\mathcal{M}'}  
				\left( C_{i}^{Non} - C_i^{O}( \boldsymbol{q}_{i}^{\ast}, \boldsymbol{x}_{n}^{\ast}, \boldsymbol{r}_{c,i}^{\ast}, \boldsymbol{r}_{d,i}^{\ast}) 
				\right) }  {M'} 
			\right]^{M'}. \label{optimalobjective}
		\end{equation}
	
	From (\ref{optimalobjective}), we conclude that Problem \textbf{NBP} maximizes the social welfare of all microgrids in $\mathcal{M}'$, \emph{i.e.}, \( \displaystyle \sum_{i\in\mathcal{M}'} \left( C_{i}^{Non} - C_i^{O}( \boldsymbol{q}_{i}^{\ast}, \boldsymbol{x}_{n}^{\ast}, \boldsymbol{r}_{c,i}^{\ast}, \boldsymbol{r}_{d,i}^{\ast}) \right)\). Since $C_{i}^{Non}$ is given, we prove that Problem \textbf{NBP} minimizes the social cost of all microgrids in $\mathcal{M}'$, \emph{i.e.}, \( \displaystyle \sum_{i\in\mathcal{M}'} C_i^{O}( \boldsymbol{q}_{i}^{\ast}, \boldsymbol{x}_{n}^{\ast}, \boldsymbol{r}_{c,i}^{\ast}, \boldsymbol{r}_{d,i}^{\ast}) \).

	\subsection{Proof of Theorem 1}
	For microgrids in set $\mathcal{M}'$, we have proved in \textbf{Proposition 1} that Problem \textbf{NBP} minimizes the social cost of all microgrids in $\mathcal{M}'$. Therefore, we can solve Problem \textbf{NBP} in a sequential manner. First, we solve the optimal energy trading \& power scheduling to minimize the social cost of microgrids in $\mathcal{M}'$. Second, we solve the optimal trading payment to maximize the Nash product in \textbf{NBP}.
	
	However, we do not know the set $\mathcal{M}'$. Since a microgrid $j \in \mathcal{M} \backslash \mathcal{M}'$ does not have incentives to trade energy, its optimal energy trading decisions will be $\boldsymbol{e}_{j}^{\ast} = \boldsymbol{0}$, and its optimal cost is the same as the benchmark cost $C_{j}^{Non}$. Therefore, we can solve the cost minimization problem for the whole system consisting of all the microgrids in set $\mathcal{M}$, and determine the set $\mathcal{M}'$ in which microgrids trade energy.
	
	Specifically, we have the following two steps to solve Problem \textbf{NBP}. In the first step, we minimize the social cost of the microgrids-system in Problem \textbf{P1}:
		\begin{equation*}
			\begin{aligned}
				& \min && \sum_{i\in\mathcal{M}} C_i^{O}( \boldsymbol{q}_{i}, \boldsymbol{x}_{n}, \boldsymbol{r}_{c,i}, \boldsymbol{r}_{d,i}) \\
				& \text{subject to} 
				&& \text{\eqref{constraint-wind}--\eqref{constraint-sold}, \eqref{constraint-load1}, \eqref{constraint-load2}, \eqref{constraint-storage1}--\eqref{constraint-storage4},  \eqref{constraint-selling}, \eqref{constraint-trading1}, and \eqref{constraint-trading3}},\\
				& \text{variables:}
				&&\{ \boldsymbol{g}_{i}, \boldsymbol{q}_{i}, \boldsymbol{x}_{n}, \boldsymbol{r}_{c,i}, \boldsymbol{r}_{d,i}, \boldsymbol{e}_{i}, ~i \in \mathcal{M} \}.
			\end{aligned} 
		\end{equation*}
	
	Solving \textbf{P1} determines the optimal energy trading \& scheduling $ \{\boldsymbol{g}_{i}^{\ast},\boldsymbol{q}_{i}^{\ast},\boldsymbol{x}_{n}^{\ast},\boldsymbol{r}_{c,i}^{\ast},\boldsymbol{r}_{d,i}^{\ast}, \boldsymbol{e}_{i}^{\ast} \}$ for each microgrid $i \in \mathcal{M}$. Based on the optimal energy trading decisions, we determine the set $\mathcal{M}'$ in the following. Specifically, set $\mathcal{M}'$ is made up of microgrids having nonzero energy trading vector $\boldsymbol{e}_{i}^{\ast}$. For each of the remaining microgrids $j \in \mathcal{M} \backslash \mathcal{M}'$, it has a zero energy trading vector, \emph{i.e.}, $\boldsymbol{e}_{j}^{\ast} = \boldsymbol{0}$, and thus neither receives nor makes payment, \emph{i.e.}, $\boldsymbol{\pi}_{j}^{\ast} = \boldsymbol{0}$. Only those microgrids in set $\mathcal{M}'$ would bargain with each other to decide the associated trading payments.
	
	In the second step, we solve the optimal trading payment for microgrids $i \in \mathcal{M}'$ in Problem \textbf{P2}:
		\begin{equation*}
			\begin{aligned}
				& \max 
				&& \prod_{i\in\mathcal{M}'} \Big( \delta_{i}^{\ast} -  C_{e} (\boldsymbol{\pi}_{i}) \Big)  \\
				& \text{subject to} 
				&& \text{\eqref{constraint-trading2} and \eqref{constraint-trading4}},\\
				& \text{variables:} 
				&& \{ \boldsymbol{\pi}_{i},~i \in \mathcal{M}' \},
			\end{aligned}
		\end{equation*}
		where \( \displaystyle \delta_{i}^{\ast} \triangleq C_{i}^{Non} - C_i^{O}( \boldsymbol{q}_{i}^{\ast}, \boldsymbol{x}_{n}^{\ast}, \boldsymbol{r}_{c,i}^{\ast}, \boldsymbol{r}_{d,i}^{\ast}) \). Solving \textbf{P2} determines the optimal payments for those microgrids participating in energy trading. 
	
\end{document}